\documentclass[12pt]{article}
\usepackage{bm}
\usepackage{a4}
\usepackage{amsthm}
\usepackage{amsmath}
\usepackage{amssymb}
\usepackage{graphicx}
\usepackage{mathtools}
\usepackage{xcolor}
\usepackage{graphics}
\usepackage{setspace}
\usepackage[shortlabels]{enumitem}
\evensidemargin \oddsidemargin
\definecolor{drab}{rgb}{0.59, 0.44, 0.09}
\definecolor{applegreen}{rgb}{0.55, 0.71, 0.0}
\definecolor{auburn}{rgb}{0.43, 0.21, 0.1}
\definecolor{awesome}{rgb}{1.0, 0.13, 0.32}
\definecolor{britishracinggreen}{rgb}{0.0, 0.26, 0.15}
\definecolor{cadmiumgreen}{rgb}{0.0, 0.42, 0.24}
\definecolor{darksalmon}{rgb}{0.91, 0.59, 0.48}

\newcommand{\CC}{\mathbb{C}}
\newcommand{\RR}{\mathbb{R}}

\newcommand{\NN}{\mathbb{N}}
\newcommand{\cupdot}{\mathbin{\mathaccent\cdot\cup}}

\newcommand{\Hi}{{\cal H}}
\newcommand{\Ob}{{\cal O}}

\newcommand{\A}{{\cal A}}
\newcommand{\B}{{\cal B}}

\newcommand{\bx}{\bm{x}}
\newcommand{\bL}{\bm{L}}

\newcommand{\mathsym}[1]{{}}
\newcommand{\unicode}[1]{{}}

\newcommand{\be}{\begin{equation}}
\newcommand{\ee}{\end{equation}}
\newcommand{\bea}{\begin{eqnarray}}
\newcommand{\eea}{\end{eqnarray}}
\newcommand{\ba}{\begin{array}}
\newcommand{\ea}{\end{array}}
\def\nn{\nonumber \\}

\newtheorem{teorema}{Theorem}[section]

\newtheorem{corollario}{Corollary}[section]

\newtheorem{lemma}{Lemma}[section]

\newtheorem{propo}{Proposition}[section]

\newtheorem{remark}{Remark}

\begin{document}
\title{The $x_i$-eigenvalue problem on some new fuzzy spheres}
\date{}

\author{Gaetano Fiore, Francesco Pisacane
   \\   \\  
Dip. di Matematica e applicazioni, Universit\`a di Napoli ``Federico II'',\\
\& INFN, Sezione di Napoli, \\
Complesso Universitario  M. S. Angelo, Via Cintia, 80126 Napoli, Italy}

\maketitle

\begin{abstract}
\noindent
We study the eigenvalue equation for the ``Cartesian coordinates'' observables $x_i$
on the fully $O(2)$-covariant fuzzy circle $\{S^1_\Lambda\}_{\Lambda\in\NN}$ ($i=1,2$) and on the fully $O(3)$-covariant fuzzy 2-sphere $\{S^2_\Lambda\}_{\Lambda\in\NN}$ ($i=1,2,3$)  introduced in 
[G. Fiore, F. Pisacane, 
J. Geom. Phys. 132 (2018), 423-451].
We show that the spectrum and eigenvectors of $x_i$ fulfill a number of properties which are expected for $x_i$  to approximate well the corresponding coordinate operator of a quantum particle forced to stay on the unit sphere.
\end{abstract}

\section{Introduction and preliminaries}
\label{introduzione}

Since their introduction fuzzy spaces  have raised a lively interest as a non-perturbative  technique in  quantum field or string theory based on a finite-discretization 
of space(time)  alternative to lattices. 
A fuzzy space is a particular type of noncommutative deformation of a space, more precisely
a sequence $\{\mathcal{A}_n\}_{n\in\NN}$
of  finite-dimensional (noncommutative) algebras such that  as $n$ diverges
 $\mathcal{A}_n$ goes to the commutative algebra $\mathcal{A}$
of regular functions (with pointwise product) on an ordinary manifold $M$ 
(in particular, the dimension of $\mathcal{A}_n$ diverges). Its main advantage
with respect to a lattice discretization is that the algebras 
$\A_n$ can carry representations of a Lie group (not only of a discrete one).
The first and seminal fuzzy space is the Fuzzy 2-Sphere (FS) $S^2_n$ of Madore and Hoppe \cite{Madore,HopdeWNic}: 
${\cal A}_n\simeq M_n(\CC)$ (the algebra of complex $n\times n$ matrices)
is generated by coordinate operators $\left\{x_i\right\}_{i=1}^{3}$ fulfilling
\be
[x_h,x_j]=\frac {2i}{\sqrt{n^2\!-\!1}}\varepsilon^{hjk}x_k, \qquad\qquad
x_hx_h=1                     \label{FS}
\ee
(here $n\in\NN\setminus \{1\}$, and sum over repeated indices is understood);
in fact these are obtained by the rescaling 
\be
x_i=\frac{2L_i}{\sqrt{n^2\!-\!1}},\quad i=1,2,3                   \label{rescale}
\ee
 of the elements
$L_i$ of  the standard basis of $so(3)$ in the 
irreducible representation $(\pi_l,V_l)$ characterized by $\bL^2:=L_iL_i=l(l\!+\!1)$,  or equivalently of dimension $n=2l\!+\!1$.
In a quantum field theory (QFT) on a fuzzy space the ``cutoff'' $n$ 
works as 
a regularizing parameter of ultraviolet divergences, because  integration over fields  amounts to integration over matrices of a finite size, growing with $n$
(see e.g. \cite{GroMad92,GroKliPre96'} for the first QFT on the FS, and \cite{GroKliPre96,Ramgoolam,Dolan:2003th,Ste17} for examples of QFT on fuzzy spheres of higher dimensions); if spacetime $M$ is enlarged to a higher-dimensional one $M'=M\times S_n$ - where $S_n$ is a fuzzy space, 
instead of a compact manifold $S$ - it reduces the number of massive Kaluza-Klein modes of a field theory on $M'$  to a finite value \cite{AscMadManSteZou}; finally, it has been recently proposed \cite{AlessioArzano} that $n$ may also parametrize the large
(but finite) amount of information hidden in a black hole.
In the matrix model formulations of  $M$-theory \cite{Banks, Berko} and string theory
\cite{IKKT97} fuzzy spaces may arise as subalgebras 
 giving the leading contribution to the path-integrals over  larger matrix algebras;
they respectively lead to quantized branes (in particular, the 5-brane) in a 11- or 10- dimensional spacetime.

Relations (\ref{FS}) are covariant under $SO(3)$, but not under the whole $O(3)$,  
in particular not under inversion of axes $x_i\mapsto -x_i$. This is to be contrasted with the $O(3)$-covariance of  the  ordinary sphere $S^2$, where the right-hand side 
of (\ref{FS})$_1$ is zero. Moreover, while the Hilbert space 
$V_l$ of the FS carries an irreducible representations of $SO(3)$,  that ${\cal L}^2(S^2)$
 of a quantum particle on $S^2$ decomposes as the direct sum of {\it all} the irreducible representations of $SO(3)$,
\be
{\cal L}^2(S^2)=\bigoplus\limits_{l=0}^\infty V_l. 
\label{directsum}
\ee
To overcome these shortcomings, in \cite{FiorePisacaneJGP18,FiorePisacanePOS18} we have built new fuzzy spheres $\{S^1_\Lambda\}_{\Lambda\in\NN}$ and $\{S^2_\Lambda\}_{\Lambda\in\NN}$, which are a fully 
$O(2)$-covariant fuzzy circle and a fully  $O(3)$-covariant fuzzy 2-sphere, respectively;
the right-hand side of (\ref{FS})$_1$ depends on the angular momentum components 
and therefore is parity invariant as in Snyder  commutation relations \cite{Snyder},
see eq. (\ref{xx})$_1$ below.
The Hilbert space on which the algebra $\A_\Lambda$
of $S^2_\Lambda$ acts decomposes as  
the direct sum 
$\mathcal{H}_{\Lambda}=\bigoplus\nolimits_{l=0}^\Lambda V_l$ \
of all the irreducible representations of $SO(3)$ up to the cutoff,  and therefore also in this aspect $S^2_\Lambda$ better approximates the configuration space $S^2$  in the limit $\Lambda\to\infty$. We have constructed these fuzzy spheres
 imposing a suitable energy cutoff on  a quantum particle   subject to a confining potential well $V(r)$  with a very sharp minimum on the sphere of  radius $r=1$ in the Euclidean spaces $\RR^2,\RR^3$ respectively; the cutoff and the sharpness of the potential well are parametrized by (and diverge with) the natural  number $\Lambda$. We think that
 these new fuzzy spheres might have  applications not only  in QFT or string theory,
but also in some condensed matter physics problem as models with an effective one- or two-dimensional configuration space in the form of a circle, a cylinder or a sphere (like nanotubes, quantum waveguides, cylindrical or spherical sheets of graphene,  etc). In 
fact, in these cases parity is respected, and the restriction to the circle, cylinder or sphere is an effective one obtained ``a posteriori'' from the exact dynamics in higher dimension.

 In the present work we start to  study the localizability on $S^d_\Lambda$,
$d=1,2$, more precisely the eigenvalue equation for the coordinates $x_i$.  
This is preparatory to a number of purposes. As we shall explain,
eigenvectors of one of the coordinates are close to the most localized  states.
Localized states, especially when arranged in systems of coherent states 
\cite{CS-FiorePisacane}, will be an extremely useful tool
 for studying path integrals (partition and correlation functions) in  QFT
over the  $S^d_\Lambda$ (as over other fuzzy spaces, see e.g. \cite{BarrettEtAl2011,Ste16NPB}), as well as the quantum metric aspects of the
 $S^d_\Lambda$, in particular 
 the ``distance'' (either the spectral distance of Connes \cite{Connes,ConCha10,DanLizMar14}, or alternative ones, see e.g. \cite{BahDopFrePia11,MarMerTom12}) between two localized states.
For $x_i$ ($i=1,...,D\!\equiv\! d\!+\!1$) to approximate well and in an $O(D)$-equivariant way the corresponding coordinate of a quantum particle forced to stay on the unit sphere $S^d$, its spectrum $\Sigma_{x_i}$ should fulfill at least the following properties, which are fulfilled also by the Madore FS:
\begin{enumerate}
\item The spectrum $\Sigma_{x_i}$ of each $x_i$, for all choices of the orthogonal axes,  is the same.
\item If $\alpha$ is an eigenvalue of $x_i$, then also $-\alpha$ is.
\item In the commutative limit the spectrum $\Sigma_{x_i}$  becomes uniformly dense in $[-1,1]$, in particular the maximal and the minimal eigenvalues  converge to $1$ and $-1$, respectively.
\end{enumerate}
We are going to show that  $\Sigma_{x_i}$ on  $S^d_\Lambda$
fulfills  these and other properties. Among the latter, one, not shared by the FS,  justifies why in our opinion 
(see section \ref{conclu})  $S^2_\Lambda$ can be interpreted  as a fuzzy {\it configuration space},  while the FS should be interpreted only as a fuzzy {\it spin phase space}: namely that the eigenstate 
of $x_3$ with maximal eigenvalue, which is very localized around the North pole of $S^2$, is an eigenstate of $L_3$ with zero eigenvalue.  We adopt \cite{CS-FiorePisacane} as a measure of the localization of a state the  square space-uncertainty (dispersion) in the ambient Euclidean space $\RR^D$ ($D=2,3$), i.e. the expectation value on the state (variance)
\be\label{uncnostra}
(\Delta \bm{x})^2: 
=\left\langle \left( \bm{x}\!-\!\left\langle\bm{x}\right\rangle\right) ^2\right\rangle
=\left\langle\bm{x}\,{}^2\right\rangle- \left\langle \bm{x}\right\rangle^2
=\sum_{i=1}^D\left\langle x_i^2\right\rangle-\sum_{i=1}^D\left\langle x_i\right\rangle^2,
\ee
which is manifestly $O(D)$-invariant. This symmetry means $(\Delta \bm{x})^2_{\psi}=(\Delta R\bm{x})^2_{\psi}$ for every state $\psi\in\mathcal{H}_{\Lambda}$ and $O(D)$-transformation $R$, and implies that 
minimizing (\ref{uncnostra}) or $\left\langle \bm{x}^2\right\rangle-\left\langle x_i\right\rangle^2$ with a fixed $i\in\left\{1,\cdots,D\right\}$ is equivalent. On the other hand, since on our fuzzy spheres $\bm{x}^2\simeq 1$, this approximately amounts to maximing $\left\langle x_i\right\rangle^2$,  what occurs on the $x_i$-eigenstate  with highest eigenvalue. 
Therefore the eigenstate $\bm{\chi}$  with maximal eigenvalue of any coordinate is very localized (almost an optimally localized state; the latter are the closest to  ``classical" states). 
Evaluating (\ref{uncnostra}) on the approximation of  $\bm{\chi}$  determined in the present work is sufficient to prove \cite{CS-FiorePisacane} the bound
$$
(\Delta \bm{x})^2_{min}< \frac{C}{(\Lambda+1)^2},
$$
where $C=3.5$ for $S^1_\Lambda$ and $C=11$ for $S^2_\Lambda$, which is 
lower than in the Madore-Hoppe FS. 

The plan of the paper is as follows. In  {section \ref{preliminari}} we briefly recall the construction procedure \cite{FiorePisacaneJGP18,FiorePisacanePOS18} of these fuzzy spaces and how to diagonalize Toeplitz tridiagonal  matrices; in  {sections \ref{CSD=2},\ref{CSD=3}} we study the $x_i$-eigenvalue equation on 
$S^1_\Lambda, S^2_\Lambda$ respectively; 
in {section \ref{conclu}} we compare results on $S^2_\Lambda$ and  FS; 
in {section \ref{app}} (the Appendix) we have concentrated lengthy calculations
and complex proofs.

\section{Preliminaries}\label{preliminari}

\subsection{Construction procedure of the fuzzy spheres $S^d_\Lambda$ in brief}\label{preliminari3}

Here we recall how quantum mechanics on the  $O(D)$-covariant  fuzzy spheres $S^d_\Lambda$ ($D=d\!+\!1$, \ $d=1,2$) has been introduced in \cite{FiorePisacaneJGP18}. 
We start with a  zero-spin quantum particle in $\RR^D$  configuration space 
with Hamiltonian
\be
H=-\frac 12\Delta + V(r)=-\frac 12\left[\partial_r^2+(D-1)\frac 1r\partial_r-\frac 1{r^2}L^2\right]+ V(r).                                                 \label{Ham}
\ee
Here $\Delta:= \partial_i\partial_i$, 
$\partial_i\equiv \partial/\partial x_i$ ($i=1,...,D$), $r^2:=\bx^2= x_ix_i$, $\partial_r:= \partial/\partial r$. We use dimensionless cartesian coordinates $x_i$, momentum components $p_i:=-i\partial_i$ and Hamiltonian  $H$; \ $x_i,p_i$ generate the Heisenberg algebra $\Ob$ of observables. Moreover $L_{ij}:=x_ip_j-x_jp_i$ are the angular momentum components, and $\bL^2:=L_{ij}L_{ij}/2$ 
is the square angular momentum (in normalized units), i.e. the Laplacian on the sphere $S^d$.
The  canonical commutation relations  
as well as $H$ are invariant under all orthogonal transformations,
including parity. 
We choose $V(r)$ as a confining potential  with a very sharp minimum at \ $r=1$, i.e. with $V'(1)=0$ and
very large $k:= V''(1)/4>0$, \ and fix 
$V_0:= V(1)$ so that the ground state has zero energy,  $E_0=0$. 
We  choose an energy cutoff $\overline{E}$ satisfying first of all the condition
\be
V(r)\simeq V_0+2k (r-1)^2\qquad \mbox{if $r$ fulfills}\quad V(r)\le  \overline{E},
\label{cond1}
\ee
so that $V(r)$ is approximately  harmonic  in the
classical region $v_{\overline{E}}$ compatible with the energy cutoff \ $V(r)\le \overline{E}$. \ 
Then we project the theory onto the finite-dimensional Hilbert subspace $\mathcal{H}_{\overline{E}}\subset\Hi\equiv \mathcal{L}^2(\RR^D)$ 
spanned by $\psi$ fulfilling the eigenvalue equation
\be
H\psi=E\psi, \qquad\psi\in {\cal L}^2\left(\mathbb{R}^D\right), \quad  E\leq\overline{E}. \label{Heigen}
\ee
This entails  replacing every observable $A$ by $\overline{A}$: 
$$
A\mapsto\overline{A}:=P_{\overline{E}}AP_{\overline{E}},
$$
where $P_{\overline{E}}$
is the projection on $\mathcal{H}_{\overline{E}}$. In particular we thus construct the fuzzy 
Cartesian coordinates $\overline{x}_i$ and angular momentum components $\overline{L_{ij}}$.
$H,L_{ij},P_{\overline{E}}$ commute, so that $\overline{H}=P_{\overline{E}}H=HP_{\overline{E}}$, 
$\overline{L_{ij}}=P_{\overline{E}}L_{ij}=L_{ij}P_{\overline{E}}$.
Replacing the Ansatz $\psi=\tilde f(r)Y(\varphi,...)$  \ ($Y$ are eigenfunctions of $\bm{L}^2$ and of the elements of a Cartan subalgebra of $so(D)$; $r,\varphi,...$ are polar coordinates) transforms the PDE \ $H\psi=E\psi$ \ into an  ODE in the unknown $\tilde f(r)$. At leading order in $1/k$ the latter is the eigenvalue equation of a $1-$dimensional harmonic oscillator, and the lowest eigenvalues  are $E_{j,n}=j\left(j+d-1\right)+2n\left(\sqrt{2k}+d-2\right)$, $j,n\in\NN_0$. Choosing a $\Lambda\in\NN$ fulfilling
$\Lambda(\Lambda+d-1)<2\left(\sqrt{2k}+d-2\right)$, and setting $ \overline{E}=\Lambda(\Lambda+d-1)$, 
we ``freeze'' all radial excitations and make the spectrum consist only of the eigenvalues $E_j\equiv E_{j,0}=j\left(j+d-1\right)$
of $\overline{H}$, which make up the lower
part of the spectrum of  $\bL^2$ - the Laplacian on $S^d$-, as wished. Correspondingly, we re-denote $\mathcal{H}_{\overline{E}},P_{\overline{E}}$ as $\mathcal{H}_{\Lambda},P_{\Lambda}$.
Consistency with
(\ref{cond1}) requires $k$ to be a   function of $\Lambda$ growing sufficiently fast with $\Lambda$, e.g.  $k(\Lambda)\ge\Lambda^2(\Lambda\!+\!1)^2$.  
Finally, we denote as $\mathcal{A}_{\Lambda}$ the algebra End$(\mathcal{H}_{\Lambda})$ of observables on $\mathcal{H}_{\Lambda}$.
Below we shall remove the bar and denote the generic operator $\overline{A}\in\mathcal{A}_{\Lambda}$   as $A$ .

The manifest $O(D)$-equivariance of $S^d_\Lambda$ has a number
of welcome consequences, in particular implies that the spectra $\Sigma_{x_i}$ 
fulfill properties 1,2 mentioned in the introduction.

\subsection{Diagonalization of Toeplitz tridiagonal  matrices}\label{preliminari4}
A real Toeplitz tri-diagonal matrix is a $n\times n$ matrix
\be
P_n\left(a,b,c\right):=\left(
\begin{array}{cccccccc}
a&b&0&0&0&0&0&0\\
c&a&b&0&0&0&0&0\\
0&c&a&b&0&0&0&0\\
\vdots&\vdots&\vdots&\vdots&\ddots&\vdots&\vdots&\vdots\\
0&0&0&0&\cdots&a&b&0\\
0&0&0&0&\cdots&c&a&b\\
0&0&0&0&\cdots&0&c&a\\
\end{array}
\right) \quad \mbox{where }a,b,c\in\mathbb{R}.                    \label{Toeplitz}
\ee
Its eigenvalues are (see e.g. \cite{Noschese} p. 2-3)
\be\label{eigenvaluesToe}
\lambda_h=a+2\sqrt{bc}\cos{\left(\frac{h\pi}{n+1}\right)},\quad h=1,\cdots,n
\ee
and  the corresponding eigenvectors $\chi^{h}$ are columns with the following components
\be\label{eigvToe}
\chi^{h,k}=\left(\frac{c}{b}\right)^{\frac{k}{2}}\sin{\left(\frac{hk\pi}{n+1}\right)},\qquad h,k=1,2,\cdots,n,
\ee
up to normalization. 
In the symmetric case ($b=c$) all eigenvalues are real and  the highest one is clearly 
$\lambda_1$; the norm of $\chi^1$ is  easily computed:
\be\label{normalization}
\left\Vert\chi^1\right\Vert^2=\sum_{k=1}^{n}\sin^2{\left(\frac{k\pi}{n+1}\right)}=\frac{n+1}{2}. 
\ee

\section{The fuzzy circle $S^1_\Lambda$}
         \label{CSD=2}

\subsection{Preliminaries}
The case $D=2$ leads to the $O(2)$-covariant fuzzy circle  $S^1_\Lambda$.  A suitable orthonormal basis $\B:=\{\psi_\Lambda,\psi_{\Lambda-1},...,\psi_{-\Lambda}\}$ of the Hilbert space $\Hi_\Lambda$ consists of eigenvectors of the angular momentum $L\equiv L_{12}$,
\be\label{azioneL1}
L \psi_n=n\psi_n.
\ee
Beside the  hermitean  cartesian coordinates $x_1,x_2$  we use the
hermitean conjugate ones\footnote{We have changed conventions with respect to  \cite{FiorePisacaneJGP18}:
the  $x_i$ ($i=1,2$) as defined here equal the $\xi^i=\overline{x}^i/a$ of \cite{FiorePisacaneJGP18}
where \ $a=1\!+\!\frac 94\frac{1}{\sqrt{2k}}\!+\!O\!\left(\!\frac{1}{k}\!\right)$ is just a normalization  factor; the $x_\pm$ as defined here equal  $\sqrt{2}\xi^\pm=\sqrt{2}\overline{x}^\pm/a$ of \cite{FiorePisacaneJGP18}.}
$$
 x_{\pm}:=x_1\pm ix_2;
$$
they act as follows:
\bea
x_+\psi_n=
 b_{n+1}\psi_{n+1} ,  \qquad x_-\psi_n=b_n\psi_{n-1} ,  \qquad 
b_n:=\left\{\!\!\ba{ll}\displaystyle    \sqrt{1\!+\!\frac{n(n \!-\! 1)}{k}} \:\: &
\mbox{if }1\!-\!\Lambda\leq  n\leq\Lambda, \\[10pt]
0 & \mbox{otherwise}.
\ea\right. 
      \label{defLxiD=2}
\eea
It is easy to see that $b_{-\Lambda}=b_{\Lambda+1}=0$, \ \ $b_n=b_{1-n}$ \ if $\Lambda+1\ge n\ge 0$.
The operator $\bx^2:=x_1 ^2+x_2^2=(x_+x_-+x_-x_+)/2$  represents the square distance from the origin.
We denote as $\widetilde{P}_m$ the projection over the $1$-dim subspace spanned by $\psi_m$.
The above formulae lead to the following $O(2)$-equivariant algebraic relations:
\be
\left[L , x_{\pm}\right]=\pm x_\pm,\quad x_+{}^\dagger=x_-, \qquad L ^\dagger=L,\label{commrelD=2'}
\ee
\be
\left[x_+,x_-\right]=-\frac{2L }k+\left[1\!+\!\frac {\Lambda(\Lambda\!+\!1)}k\right]\!\left(\widetilde P_{\Lambda}\!-\!\widetilde P_{-\Lambda}\right),\label{y+y-}
\ee
\be
\bx^2=  1+\frac{L^2}{k} -
\left[1\!+\!\frac {\Lambda(\Lambda\!+\!1)}{k}\right]
\frac{\widetilde P_{\Lambda}\!+\!\widetilde P_{-\Lambda}}2,          \label{defR2D=2}
\ee
\be
\prod\limits_{m=-\Lambda}^{\Lambda}\!\!\left(L \!-\!mI\right)=0, \qquad \left(x_\pm\right)^{2\Lambda+1}=0 .\label{commrelD=2}
\ee
Formula (\ref{defR2D=2}) shows that $\bx^2$ is not the identity, but  a function of $L^2$, hence the $\psi_m$ are
its eigenvectors; its eigenvalues  
(except on $\psi_{\pm\Lambda}$)
are close to 1, slightly grow with $|m|$ and collapse to 1 as $\Lambda\to \infty$.

Finally, we showed that there is a sequence of $O(2)$-covariant $*$-algebra isomorphisms 
$\A_{\Lambda}\simeq M_N(\CC)\simeq\pi_\Lambda[Uso(3)]$, where $\pi_{\Lambda}$ is the 
$(2\Lambda\!+\!1)$-dimensional unitary irreducible representation  of $Uso(3)$.

\subsection{Spectrum of $x_i$ in the $O(2)$-equivariant fuzzy circle}
\label{circlespec}
In this subsection we analyze the spectrum of $x_1$. This is not a restriction because the algebraic relations (\ref{commrelD=2'}-\ref{commrelD=2}) are covariant under $O(2)$ transformations $\bm{x}\mapsto \bm{x}'=R\bm{x}$, $L$ is covariant under $2$-dimensional rotations, $L\to -L$ under $x_1$-inversion and the same applies under $x_2$-inversion; this implies that the spectra $\Sigma_{x_i}\left(\Lambda\right)$ of all coordinate operators $x_i$ are equal, and for this reason we can focus our attention to $x_1$. The spectrum $\Sigma_{x_1}$ for $\Lambda=1,2$ is presented in 
formulae (\ref{eigenv-x_1S^1_1}-\ref{eigenv-x_1S^1_2}) of the appendix.

More generally, on the basis $\B$ of $\Hi_\Lambda$  the operator 
$x_1$ is represented by the $(2\Lambda\!+\!1)\times(2\Lambda\!+\!1)$ symmetric tri-diagonal matrix  [cf. (\ref{Toeplitz})]
$$
X^\Lambda=\frac 12\left(
\begin{array}{cccccccc}
0&b_{\Lambda} &0&0&0&0&0&0\\
b_{\Lambda}  &0&b_{\Lambda-1} &0&0&0&0&0\\
0&b_{\Lambda-1} & 0&b_{\Lambda-2} &0&0&0&0\\
\vdots&\vdots&\vdots&\vdots&\ddots&\vdots&\vdots&\vdots\\
0&0&0&0&\cdots&b_{2-\Lambda} &0&b_{1-\Lambda} \\
0&0&0&0&\cdots&0&b_{1-\Lambda} &0 
\end{array}
\right)         =X^\Lambda_0+O\!\left(\!\frac {1}{\Lambda^2}\!\right),
$$
where $X_0^\Lambda:=\frac 12 P(0,1,1) $, and it is obvious that all the eigenvalues of $X^{\Lambda}$ are real.

Let $\Sigma^\Lambda_0:=\left\{\widetilde{\alpha}_h(\Lambda)\right\}_{h=1}^{2\Lambda+1}$ be the set of the eigenvalues of $X^{\Lambda}_0$ arranged in descending order; according to (\ref{eigenvaluesToe}) one has
\be\label{valuecos}
\widetilde{\alpha}_h(\Lambda)=\cos{\left(\frac{h \pi}{2\Lambda+2} \right)},\quad h=1,2,\cdots,2\Lambda+1.
\ee
It is easy to see that $\alpha\in \Sigma^\Lambda_0\Rightarrow -\alpha\in \Sigma^\Lambda_0$, all the eigenvalues of $\Sigma^{\Lambda}_0$ are simple, $\widetilde{\alpha}_1(\Lambda+1)>\widetilde{\alpha}_1(\Lambda)$ and $\Sigma^\Lambda_0$ becomes uniformly dense in $[-1,1]$ as $\Lambda\to \infty$.

In {section \ref{proofsD=2}} we show that the same holds true also for the spectrum $\Sigma^{\Lambda}$ of $X^\Lambda$, in particular we prove

\begin{teorema}
\label{teospec1}
\begin{enumerate}[(A)]
\item If $\alpha$ is an eigenvalue of $X^{\Lambda}$, then also $-\alpha$ is.
\item For all $\Lambda$, all  eigenvalues of $X^{\Lambda}$ are simple; we denote them as \ $\alpha_1(\Lambda),
\alpha_2(\Lambda),$ $...,\alpha_{2\Lambda+1}(\Lambda)$, \ in decreasing order.
\item Let $k(\Lambda)\geq \Lambda(\Lambda-1)(2\Lambda+3)^2(2\Lambda+4)^4/4\pi^4$, then
\be
\alpha_1\left(\Lambda+1\right)>\alpha_1\left(\Lambda\right)\quad\forall\Lambda\in\mathbb{N}.
\ee
\item $\Sigma^{\Lambda}$ becomes uniformly dense in $[-1,1]$ as $\Lambda\to \infty$, in particular
\be\label{limitvaluesalpha1}
\lim_{\Lambda\rightarrow+\infty}\alpha_1\left(\Lambda\right)=1\quad \mbox{and}\quad \alpha_1\left(\Lambda\right)\geq 1-\frac{\pi^2}{8(\Lambda+1)^2}\quad \forall\Lambda\in\mathbb{N}.
\ee
\end{enumerate}
\end{teorema}
Let $\bm{\chi}:=\sum_{n=-\Lambda}^{\Lambda}{\chi_n\psi_n}$, the eigenvalue equation $x_1\bm{\chi}=\alpha\bm{\chi}$
amounts to
\bea
\frac{b_{\Lambda}}{2}\chi_{\pm (\Lambda-1)}=\alpha\chi_{\pm \Lambda}, \qquad \qquad
\frac{b_n\chi_{n-1}+b_{n+1}\chi_{n+1}}{2}=\alpha\chi_n \qquad \mbox{if }\: |n|<\Lambda;
\eea
on the other hand, $b_n\rightarrow 1$ in the commutative limit and in {section \ref{proofteodense1}} we show that $\alpha_h\left(\Lambda\right)\simeq \cos{\left(\frac{h\pi}{2\Lambda+2}\right)}$ in the limit $\Lambda\rightarrow+\infty$, so (\ref{eigvToe}) and (\ref{normalization}) imply
$$
x_1 \bm{\chi}_h\left(\Lambda\right)=\alpha_h\left(\Lambda\right)\bm{\chi}_h\left(\Lambda\right)\Longrightarrow \chi_{h,n}\left(\Lambda\right)\simeq \sqrt{\frac{2}{2\Lambda+2}}\sin{\left( \frac{hn\pi}{2\Lambda+2}\right)}.
$$

\section{The fuzzy sphere $S^2_\Lambda$}
  \label{CSD=3}

\subsection{Preliminaries}\label{preliminari2}
The case $D=3$ leads to the $O(3)$-covariant fuzzy sphere  $S^2_\Lambda$.  A suitable orthonormal basis $\B:=\{\psi_l^m:l,m\in\mathbb{Z}, |m|\leq l\leq \Lambda\}$ of the Hilbert space $\Hi_\Lambda$ consists of common eigenvectors of the angular momentum component $L_3\equiv L_{12}$ and of the square angular momentum operator $\bL^2$,
\be\label{azioneLD=3}
L_3 \psi_l^m=m\psi_l^m,\quad\quad\quad \bL^2\psi_l^m=l(l+1)\psi_l^m.
\ee
We define $x_0:= x_3$, $L_0:=L_3$ and beside the  hermitean  cartesian coordinates $x_1,x_2$ and angular momentum components $L_1,L_2$ we use the
hermitean conjugate ones
$$
 x_{\pm}:=\frac{x_1\pm ix_2}{\sqrt{2}},\quad\quad\quad L_{\pm}:=\frac{L_1\pm iL_2}{\sqrt{2}};
$$
they act as follows (here $a\in\{0,+,-\}$):
\be\label{azioneL}
L_{\pm}{\psi}_l^m=\frac{\sqrt{(l\!\mp\! m)(l\!\pm\! m\!+\!1)}}{\sqrt{2}}{\psi}_l^{m\pm 1}\!,\quad x_a\bm{\psi}_{l}^m=\left\{\!\!
\ba{ll}
c_l A_{l}^{a,m}\bm{\psi}_{l-1}^{m+a}+
c_{l+1} B_{l}^{a,m} \bm{\psi}_{l+1}^{m+a
}&\mbox{ if }l<\Lambda,\\[8pt]
c_lA_l^{a,m}\bm{\psi}_{\Lambda-1}^{m+a}&\mbox{ if }
l=\Lambda,\\[8pt]
0&\mbox{otherwise,}
\ea
\right. 
\ee
where 
\be 
\label{Clebsch}
A_l^{0,m}=\sqrt{\frac{(l+m)(l-m)}{(2l+1)(2l-1)}}\hspace{0.3cm},\hspace{0.3cm} A_l^{\pm,m}=\frac{\pm1}{\sqrt{2}}\sqrt{\frac{(l\mp m)(l\mp m-1)}{(2l-1)(2l+1)}}\hspace{0.3cm},\hspace{0.3cm} A_l^{a,m}=B_{l-1}^{-a,m-a},
\ee
\be\label{defcl}
c_l:= \sqrt{1+\frac{l^2}{k}}\qquad 1\le l\le \Lambda,\qquad c_0=c_{\Lambda+1}=0,\quad\mbox{with}\quad k=k\left(\Lambda\right)\geq \Lambda^2\left(\Lambda+1\right)^2.
\ee
The choice (\ref{defcl})$_1$ is compatible with all $V(r)$ having the same $V(1)$, $V'(1)=0$ and $V''(1)=4k$, up to $O\left(\frac{1}{k^{\frac{3}{2}}}\right)$.

The operator $\bx^2:=x_1 ^2+x_2^2+x_3^2=x_+x_-+x_-x_++x_3^2$  represents the square distance from the origin.
We denote as $\widetilde{P}_l$ the projection over the $\bL^2\equiv l(l+1)$ eigenspace. The above formulae lead to the following $O(3)$-equivariant algebraic relations:

\be
[L_i,x_j]=i\varepsilon^{ijh}x_h, \qquad 
\left[\,L_i,L_j\right]=i\varepsilon^{ijh}L_h,\quad x_{i}^{\dag}=x_i, \qquad 
L_i ^{\dag}=L_i, \label{rf3D4}
\ee
\be 
x_iL_i=0,\hspace{1.5cm}
[x_i,x_j]=i\varepsilon^{ijh}\left(-\frac{I}{k}+K\widetilde{P}_{\Lambda}\right)L_h \hspace{1.5cm}i,j,h\in\{1,2,3\}\label{xx},
\ee
\be\label{defR2D=3}
\bx^2= 1
+\frac{\bL^2+1}{k}-\left[1+\frac{(\Lambda\!+\!1)^2}{k}\right]\frac{\Lambda\!+\!1}{2\Lambda+1}\widetilde P_{\Lambda},
\ee
\be 
\prod_{l=0}^{\Lambda}\left[\bL^2-l(l+1)I\right] =0,\qquad
\prod_{m=-l}^{l}{\left(L_3-mI\right)}\widetilde{P}_l=0,\qquad \left(x_{\pm}\right)^{2\Lambda+1}=0. \label{rf3D3}
\ee
Formula (\ref{defR2D=3}) shows that $\bx^2$ is not the identity, but  a function of $\bL^2$, hence the $\psi_l^m$ are
its eigenvectors; its eigenvalues  
(except on $\psi_{\Lambda}^m$)
are close to 1, slightly grow with $l$ and collapse to 1 as $\Lambda\to \infty$.

Finally, we showed that there is a sequence of $O(3)$-covariant $*$-algebra isomorphisms 
$\A_{\Lambda}\simeq M_N(\CC)\simeq\pi_\Lambda[Uso(3)]$, where $\pi_{\Lambda}$ is the 
$(\Lambda\!+\!1)^2$-dimensional unitary irreducible representation  of $Uso(4)$.
\subsection{Spectrum of $x_i$ in the $O(3)$-equivariant fuzzy sphere}\label{spherespec}
In this subsection we do the analysis of  the spectrum of $x_0$, this is not a restriction since the covariance of the algebra under  $O(3)$ transformations $\bm{x}\mapsto \bm{x}'=R\bm{x}$, $\bm{L}\mapsto \bm{L}'=R\bm{L}$ implies that the spectra $\Sigma_{x_i}\left(\Lambda\right)$ of all coordinate operators $x_i$ of our fuzzy space are equal; on the other hand, because of $\left[x_0,L_0\right]=0$, we can simultaneously diagonalize $x_0$ and $L_0$.

Eq. (\ref{azioneLD=3})$_1$ and
\be \label{diagxL}
\begin{cases}
L_0\bm{\chi}_{\alpha}^{\beta}=\beta \bm{\chi}_{\alpha}^{\beta}\\
x_0\bm{\chi}_{\alpha}^{\beta}=\alpha \bm{\chi}_{\alpha}^{\beta}
\end{cases},
\ee
imply
\be\label{nuovacond}
\beta=m\in\{-\Lambda,-\Lambda+1,\cdots,\Lambda-1,\Lambda\}\quad\mbox{and}\quad\bm{\chi}_{\alpha}^m=\sum_{l=|m|}^{\Lambda}{\chi_{\alpha,l}^m\bm{\psi}_l^m};
\ee
so $x_0\bm{\chi}_{\alpha}^{m}=\alpha \bm{\chi}_{\alpha}^{m}
$ can be re-written as
\be
\left\{
\begin{split}
\chi_{\alpha,|m|+1}^{m}c_{|m|+1}A_{|m|+1}^{0,m}&=\alpha \chi_{\alpha,|m|}^{m}\\
\chi_{\alpha,|m|}^{m}c_{|m|+1}B_{|m|}^{0,m}+\chi_{\alpha,|m|+2}^{m}c_{|m|+2}A_{|m|+2}^{0,m}&=\alpha \chi_{\alpha,|m|+1}^{m}\\
\chi_{\alpha,|m|+1}^{m}c_{|m|+2}B_{|m|+1}^{0,m}+\chi_{\alpha,|m|+3}^{m}c_{|m|+3}A_{|m|+3}^{0,m}&=\alpha \chi_{\alpha,|m|+2}^{m}\\
\quad\vdots\quad\quad\vdots\quad\quad\vdots\quad\quad\vdots\quad\quad\vdots\quad\quad\vdots\quad\quad&\vdots\quad\quad\vdots\\
\chi_{\alpha,\Lambda-2}^{m}c_{\Lambda-1}B_{\Lambda-2}^{0,m}+\chi_{\alpha,\Lambda}^{m}c_{\Lambda}A_{\Lambda}^{0,m}&=\alpha \chi_{\alpha,\Lambda-1}^{m}\\
c_{\Lambda}B_{\Lambda-1}^{0,m}\chi_{\alpha,\Lambda-1}^{m}&=\alpha \chi_{\alpha,\Lambda}^{m}
\end{split}
\right.
\label{autovalvec}
\ee
which in turn can be rewritten in the matrix form $B_m(\Lambda)\chi=\alpha \chi$,
where $\chi=\left(\chi_{\alpha,|m|}^{m},\chi_{\alpha,|m|+1}^{m},\ldots,\chi_{\alpha,\Lambda}^{m} \right)^T$ and $B_m(\Lambda)$  is the following $n(\Lambda;m)\times n(\Lambda;m)$ symmetric tridiagonal matrix
$$
\hspace{-0.25cm}
B_m(\Lambda)=\left(\!\!
\begin{array}{cccccccc}
0&\!c_{|m|+1}A_{|m|+1}^{0,m}\!&0&0&0&0&0&0\\
c_{|m|+1}A_{|m|+1}^{0,m}\!&0&\!c_{|m|+2}A_{|m|+2}^{0,m}\!&0&0&0&0&0\\
0&\! c_{|m|+2}A_{|m|+2}^{0,m}\!&0& \!c_{|m|+3}A_{|m|+3}^{0,m}\!&0&0&0&0\\
\vdots&\vdots&\vdots&\vdots&\vdots&\vdots&\vdots&\vdots\\
0&0&0&0&0& \!c_{\Lambda-1}A_{\Lambda-1}^{0,m}\! &0&\! c_{\Lambda}A_{\Lambda}^{0,m}\\
0&0&0&0&0&0&\! c_{\Lambda}A_{\Lambda}^{0,m}\! &0\\ 
\end{array}
\!\!\right),
$$
or equivalently $M_m(\Lambda;\alpha)\,\chi=0$, where $0$ here is the null vector, and we have abbreviated
$$
n=n(\Lambda;m):=\Lambda-|m|+1,\qquad\qquad M_m(\Lambda;\alpha):=B_m(\Lambda)-\alpha I_{n(\Lambda;m)}.
$$
It is well known that  the problem of determining analytically the eigenvalues of a square matrix of large rank is absolutely not trivial, but the $B_m(\Lambda)$ have several good properties (for example they are symmetric and tri-diagonal) which will help us in studying their spectra. We start with the following

\begin{remark}
All the eigenvalues of $B_m\left(\Lambda\right)$ are real, and $B_m\left(\Lambda\right)\equiv B_{-m}\left(\Lambda\right)$ implies that we can restrict our attention to the cases $\beta=m\in\{0,1,\cdots,\Lambda\}$.\label{remark}
\end{remark}
As for the fuzzy circle, we prove
\begin{teorema}
\label{teospec2}
\begin{enumerate}[(A)]
\item If $\alpha$ is an eigenvalue of $B_m\left(\Lambda\right)$, then also $-\alpha$ is.
\item For all $\Lambda,m$, all  eigenvalues of $B_m\left(\Lambda\right)$ are simple; we denote them as \ $\alpha_1(\Lambda;m),
\alpha_2(\Lambda;m),$ $...,\alpha_{n(\Lambda;m)}(\Lambda;m)$, \ in decreasing order.
\item Let $\alpha_1\left(\Lambda;m\right)$ be the highest eigenvalue of $B_m\left(\Lambda\right)$, then
\be\label{disalpha}
\alpha_1\left(\Lambda;0\right)>\alpha_1\left(\Lambda;1\right)>\cdots>\alpha_1\left(\Lambda;\Lambda\right),
\ee
and
\be\label{disalpha1}
\alpha_1\left(\Lambda+1;0\right)>\alpha_1\left(\Lambda;0\right)\quad\mbox{definitively, if }k(\Lambda)\geq \Lambda^6.
\ee
\item $\Sigma_{B_0\left(\Lambda\right)}$ becomes uniformly dense in $[-1,1]$ as $\Lambda\to \infty$, in particular
\be \label{valuelimalpha2}
\lim_{\Lambda\rightarrow+\infty}\alpha_1\left(\Lambda;0\right)=1\quad \mbox{and}\quad \alpha_1\left(\Lambda;0\right)\geq 1-\frac{\pi^2}{2(\Lambda+2)^2} \quad \forall\Lambda\geq 2.
\ee
\end{enumerate}
\end{teorema}
Item \emph{(C)} of last theorem allows also us to make a connection between our localized states and the classical ones because the $\alpha_1\left(\Lambda;0\right)$-eigenstate approximates a quantum particle on $S^2$ concentrated (because of the above equivalence between the $\alpha_1\left(\Lambda;0\right)$-eigenstate and the most localized state of our fuzzy space \cite{CS-FiorePisacane}) on the North pole and rotating around the $x_3$-axis; on the other hand, if we consider a classical particle forced to stay on $S^2$ and in the position $(0,0,1)$, then it must be
$$\vspace{-0.1cm}L_3=\left(\underline{\bm{ L}}\right)_3=\left(\underline{\bm{r}} \times \underline{\bm{p}}\right)_3 =0,$$
as for our case.

Note that, the spectrum $\Sigma_{B_0\left(\Lambda\right)}$ contains exactly $\Lambda+1$ eigenvalues and the highest one fulfills (\ref{disalpha}), for this reason we focus our attention only on that matrix.

It is important to point out that the proof of item \emph{(D)} can be trivially re-arranged in order to prove that it holds for $\Sigma_{B_m\left(\Lambda\right)}$ and $\alpha_1\left(\Lambda;m\right)$ also if $m>0$ is any other fixed integer.

Let $m\in\mathbb{N}_0$ and assume that $\bm{\chi}_{\alpha}^m:=\sum_{l=m}^{\Lambda}\chi_{\alpha,l}^m \bm{\psi}_l^m$ is a common eigenstate of $x_0$ and $L_0$; let $\left\{\widetilde{\alpha}_h\left(\Lambda;m\right) \right\}_{h=1}^{\Lambda-m+1}$ be the set of the eigenvalues of $P_{\Lambda-m+1}\left(0,\frac{1}{2},\frac{1}{2}\right)$ arranged in descending order; according to (\ref{eigenvaluesToe}) one has
$$
\widetilde{\alpha}_h(\Lambda,m)=\cos{\left(\frac{h \pi}{\Lambda-m+2} \right)},\quad h=1,2,\cdots,\Lambda-m+1.
$$

We can prove (as for {section \ref{proofteodense1}}) that $\alpha_h\left(\Lambda;m\right)\simeq \cos{\left(\frac{h\pi}{\Lambda-m+2}\right)}$ in the limit $\Lambda\rightarrow+\infty$, although in this case $c_lA_l^{0,m}\nrightarrow\frac{1}{2}$. On the other hand, when $|m|\ll l$, we can approximate well $c_l A_l^{0,m}\simeq\frac{1}{2}$ in the commutative limit, for this reason we believe that
$$
\chi_{\alpha_h\left(\Lambda;m\right),l}^m\simeq \sqrt{\frac{2}{\Lambda-m+2}}\sin{\left( \frac{hl\pi}{\Lambda-m+2}\right)},
$$
as for the $D=2$ case.

\section{Conclusions and comparison with the Madore fuzzy sphere}\label{conclu}

In the analysis of the spectra $\Sigma_{x_i}\left(\Lambda\right)$ of our fuzzy spaces we have proved the following:
\begin{enumerate}
\item $O(D)$-equivariance: the spectrum $\Sigma_{x_i}$ of each $x_i$, for all choices of the orthogonal axes,  is the same.
\item Parity property:
$$
\alpha\in\Sigma_{x_i}\left(\Lambda\right)\Rightarrow -\alpha\in\Sigma_{x_i}\left(\Lambda\right). 
$$
\item Monotonicity of the maximal eigenvalue with respect to $\Lambda$:
$$
\max{\Sigma_{x_i}\left(\Lambda\right)}<\max{\Sigma_{x_i}\left(\Lambda+1\right)}\quad\mbox{and}\quad\lim_{\Lambda\rightarrow+\infty}\left[\max{\Sigma_{x_i}\left(\Lambda\right)}\right]=1.
$$
\item Density property
$$
\Sigma_{x_i}\left(\Lambda\right)\mbox{ becomes uniformly dense in }[-1,1]\mbox{ when }\Lambda\rightarrow+\infty.
$$
\item On our fuzzy sphere $S^2_\Lambda$
 the state $\bm{\chi}$ most localized around  the North pole  
fulfills the property $L_3\bm{\chi}=0$ (item \emph{(C)} of {theorem \ref{teospec2}}),  as the 
generalized quantum state (distribution) $2\delta(\theta)/\sin\theta\simeq\delta(x_1)\delta(x_2)$ on $S^2$ concentrated on the North pole (here $\theta$ is the colatitude); the classical counterpart of this property is that the classical particle on $S^2$ in the position $(0,0,1)$ has zero $L_3$ ($z$-component of the angular momentum).
\end{enumerate}
It is important to underline that these are welcome properties for a  $x_i$-operator which is required to approximate well, in the commutative limit, the $x_i$-coordinate of a quantum particle forced to stay on the unit sphere $S^d$.

Moreover, the spectrum of $L_i$ is $\Sigma_{L_i}\left(\Lambda\right)=\{-\Lambda,1\!-\!\Lambda,...,\Lambda\}$ for all $i=1,2,3$, by the $SO(3)$-covariance,  and  fulfills properties 1,2 (the multiplicity
of the eigenvalue $m$ is $\Lambda\!-\!|m|\!+\!1$).

\medskip
In the Madore fuzzy sphere, since the $x_i$ are obtained by the rescaling (\ref{rescale}) of angular momentum operators acting in an irreducible representation, then all $x_i$ have again the same spectrum as $x_3$, by  $SO(3)$-covariance,  and this is obtained by the rescaling of the spectrum of  $L_3$; this leads to the  eigenvalues (all simple) and eigenvectors
$$
x_3\varphi_m=\frac{m}{\sqrt{\Lambda^2+\Lambda}}\varphi_m\qquad\mbox{with }m\in\left\{-\Lambda,\cdots,\Lambda \right\},
$$ 
where we have set $\Lambda\!\equiv\!(n\!-\!1)/2$.
Hence also in this case properties 1-4   are fulfilled. However, for this reason  there is no longer room for independent observables playing the role of angular momentum operators  on the carrier Hilbert space $V_\Lambda$, and property 5 is lost.

For this reason, and the other ones mentioned in the introduction,  from our point of view it is more natural to interpret the $L_i$ in the irreducible
representation $(\pi_\Lambda,V_\Lambda)$  still as the inthrinsic angular momentum components  of a particle
of spin $\Lambda$, and the states (rays) in $V_\Lambda$ as states on the
corresponding spin phase manifold. Then, since the spin degrees of freedom have no classical limit, it is not possible to define also position observables or
 see any state $\bm{\varphi}\in V_\Lambda$ as an approximation of a classical point in $S^2$-configuration space; the algebra $\mathcal{A}_n$ should be seen simply as the spin phase space algebra, not as a fuzzyfication of the algebra of configuration space 
observables on $S^2$.

\section{Appendix}
\label{app}

\subsection{A very useful proposition}
In the next proofs we often use the following
\begin{propo}\label{propoexistv+}
Let $A=(a_{i,j})_{i,j=1}^n$ be a square matrix such that $a_{i,j}\geq 0$ $\forall i,j$, then there exist a vector $\widehat{\chi}\in\mathbb{R}^n_+$ fulfilling
$$
\|\widehat{\chi}\|_2=1\quad\mbox{and}\quad\|A\left[\widehat{\chi}\right]\|_2=\|A\|_2.
$$
\proof
By definition
$$
\|A\|_2=
\sup_{\|\chi\|_2=1}{\|A[\chi]\|_2},
$$
the Weierstrass theorem implies that
\be\label{condizionesup}
\sup_{\|\chi\|_2=1}{\|A[\chi]\|_2}
=
\max_{\|\chi\|_2=1}{\|A[\chi]\|_2},
\ee
so we can consider a vector $\widetilde{\chi}\in\mathbb{R}^{n}$ fulfilling (\ref{condizionesup}) and $\|\widetilde{\chi}\|_2=1$. We need to prove that $\widetilde{\chi}_i\geq0$ for all $i$. If we suppose that $\widetilde{\chi}_j<0$ for some $j$ in $\{1,2,\cdots,n\}$, then we can define the vector $\widehat{\chi}:=\left(|\widetilde{\chi}_1|,|\widetilde{\chi}_2|,\cdots,|\widetilde{\chi}_{n}|\right)^T$. It is such that $\|\widehat{\chi}\|_2=\|\widetilde{\chi}\|_2=1$ and
$$
\left\|A\left[\widetilde{\chi}\right]\right\|_2=\sqrt{\sum_{i=1}^n\left(\sum_{j=1}^n a_{i,j}\widetilde{\chi}_j \right)^2}\overset{a_{i,j}\geq 0 }\leq \sqrt{\sum_{i=1}^n\left(\sum_{j=1}^n a_{i,j}|\widetilde{\chi}_j| \right)^2}= \sqrt{\sum_{i=1}^n\left(\sum_{j=1}^n a_{i,j}\widehat{\chi}_j \right)^2}= \left\|A\left[\widehat{\chi}\right]\right\|_2.
$$
This last inequality proves that we can consider in the realization of the maximum the ``positive'' vector $\widehat{\chi}$, instead of $\widetilde{\chi}$, so the proof is finished.
\endproof
\end{propo}
\begin{propo}\label{propoineqn2}
Let $A=(a_{i,j})_{i,j=1}^n$ and $B=(b_{i,j})_{i,j=1}^n$ be square matrices fulfilling $0\leq a_{i,j}\leq b_{i,j}$ $\forall i,j$, then
$$
\|A\|_2\leq \|B\|_2.
$$
\proof
According to {proposition \ref{propoexistv+}} we can consider a vector $\widehat{\chi}\in\mathbb{R}^n_+$ with $\|\widehat{\chi}\|_2=1$
fulfilling $\|A\|_2=\left\|A\left[\widehat{\chi}\right]\right\|_2$; so
$$
\|A\|_2= \sqrt{\sum_{i=1}^n\left(\sum_{j=1}^n a_{i,j}\widehat{\chi}_j \right)^2}\overset{a_{i,j}\leq b_{i,j}}\leq \sqrt{\sum_{i=1}^n\left(\sum_{j=1}^n b_{i,j}\widehat{\chi}_j \right)^2}\leq \|B\|_2.
$$
\endproof
\end{propo}

\subsection{The proofs of theorems of section \ref{circlespec}}\label{proofsD=2}
\subsubsection{Proof of item $(A)$ in theorem \ref{teospec1}}
Consider the unitary and involutive operator $U_1=U_1^\dagger=U_1^{-1}$
corresponding to the inversion operator of the $x_1$-axis (this exists
by the $O(2)$-covariance of our model\footnote{$U_1$ is obtained by projection on
$\Hi_\Lambda$ of the original unitary operator $\tilde U_1$ acting on $\mathcal{L}^2\left(\mathbb{R}^2\right)$ as follows: \ $\tilde U:\psi(x_1,x_2)\rightarrow \psi(-x_1,x_2)$.}: $U_1\,x_1\,U_1=-x_1$,
$U_1x_2 U_1=x_2$. Then $x_1\bm{\chi}=\alpha \bm{\chi}$ implies
 $x_1(U_1\bm{\chi})=-\alpha (U_1\bm{\chi})$, i.e. $U_1\bm{\chi}$ is
an eigenvector of $x_1$ with the opposite eigenvalue.

\subsubsection{Proof of item $(B)$ in theorem \ref{teospec1}}\label{proofteosimple1}

According to the last proof, if $I_n$ is the $n\times n$ identity matrix and $M_{\Lambda}\left(\alpha\right):=X^\Lambda+\alpha I_{2\Lambda+1}$, then the eigenvalue problem for $X^\Lambda$ is equivalent to solve $\det{\left[M_{\Lambda}(\alpha)\right]}=0$. In order to do this we define $M_{\Lambda}^n$ as the $n\times n$ submatrix of $M_{\Lambda}$ formed by the first $n$ rows and columns, then
$$p_{\Lambda}(\alpha):=\det{\left[M_{\Lambda}(\alpha)\right]}\quad\mbox { and }\quad p_{\Lambda}^n(\alpha):=\det\left\{M_{\Lambda}^n\left(\alpha\right)\right\}.$$ It is not difficult to see that
\begin{itemize}
\item when $\Lambda=1$, then 
\bea
\left|
\begin{array}{ccc}
\alpha&\frac{b_{1}}{2}&0\\
\frac{b_1}{2}&\alpha&\frac{b_0}{2}\\
0&\frac{b_0}{2}&\alpha\\
\end{array}
\right|=\alpha\left[\alpha^2\!-\!\frac{(b_0)^2}{4}\!-\!\frac{(b_1)^2}{4} \right]=:p_1\!\left(\alpha\right)\Longrightarrow \begin{cases}
\alpha_1(1)=\frac{\sqrt{(b_0)^2+(b_1)^2}}{2}=\frac{\sqrt{2}}{2},\\
\alpha_2(1)=0,\\
\alpha_3(1)=-\frac{\sqrt{(b_0)^2+(b_1)^2}}{2}-\frac{\sqrt{2}}{2};
\end{cases} \label{eigenv-x_1S^1_1}
\eea
\item when $\Lambda=2$, then
\bea
&&p_2(\alpha):=\left|
\begin{array}{ccccc}
\alpha&\frac{b_{2}}{2}&0&0&0\\
\frac{b_2}{2}&\alpha&\frac{b_1}{2}&0&0\\
0&\frac{b_1}{2}&\alpha&\frac{b_{0}}{2}&0\\
0&0&\frac{b_0}{2}&\alpha&\frac{b_{-1}}{2}\\
0&0&0&\frac{b_{-1}}{2}&\alpha
\end{array}
\right|\nn
&&=\alpha\left\{\alpha^4-\alpha^2\frac{\left[\left(b_{2}\right)^2+\left(b_{1}\right)^2+\left(b_{0}\right)^2+\left(b_{-1}\right)^2 \right]}{4}+\frac{\left(b_{1}b_{-1}\right)^2+\left(b_2 b_0\right)^2+\left(b_{2}b_{-1}\right)^2}{16} \right\}\nn
&&\Longrightarrow \qquad\begin{cases}
\alpha_1(2)= \sqrt{\frac{1}{8}}\sqrt{A_2+\sqrt{B_2}}=\frac{1}{2}\sqrt{3+\frac{2}{k}}, \\
\alpha_2(2)= \sqrt{\frac{1}{8}}\sqrt{A_2-\sqrt{B_2}}=\frac{1}{2}\sqrt{1+\frac{2}{k}},\\
\alpha_3(2)=0, \\
\alpha_4(2)=-\sqrt{\frac{1}{8}}\sqrt{A_2-\sqrt{B_2}}=-\frac{1}{2}\sqrt{1+\frac{2}{k}}, \\
\alpha_5(2)=-\sqrt{\frac{1}{8}}\sqrt{A_2+\sqrt{B_2}}=-\frac{1}{2}\sqrt{3+\frac{2}{k}},
\end{cases}
\label{eigenv-x_1S^1_2}
\eea
because $A_2:=\left(b_{2}\right)^2+\left(b_{1}\right)^2+\left(b_{0}\right)^2+\left(b_{-1}\right)^2=4\left(1+\frac{1}{k}\right)$ and 
$$B_2:=2\left[\left(b_{1}b_0\right)^2-\left(b_{2}b_0\right)^2+\left(b_{-1}b_0\right)^2+\left(b_{2}b_1\right)^2-\left(b_{-1}b_1\right)^2-\left(b_{2}b_{-1}\right)^2\right]+\left(b_{2}\right)^4+\left(b_{1}\right)^4+\left(b_{0}\right)^4+\left(b_{-1}\right)^4=4.$$
\item in general, when $\Lambda>2$, one can calculate $p_{\Lambda}\left(\alpha\right)$ through the use of this recursion formula:
\end{itemize}
\be
\begin{array}{c}
p_{\Lambda}^2\left(\alpha\right):=\det{\left\{M_{\Lambda}^2\left(\alpha\right)\right\}}=\alpha^2-\left(\frac{b_{\Lambda}}{2} \right)^2,\\
p_{\Lambda}^3\left(\alpha\right):=\det{\left\{M_{\Lambda}^3\left(\alpha\right)\right\}}=\alpha\left[\alpha^2-\frac{\left(b_{\Lambda} \right)^2+\left(b_{\Lambda-1} \right)^2}{4} \right],\\
p_{\Lambda}^4\left(\alpha\right):=\alpha \left[p_{\Lambda}^3\left(\alpha\right)\right]-\left(\frac{b_{\Lambda-2}}{2} \right)^2p_{\Lambda}^2\left(\alpha\right),\\
p_{\Lambda}^5\left(\alpha\right):=\alpha \left[p_{\Lambda}^4\left(\alpha\right)\right]-\left(\frac{b_{\Lambda-3}}{2} \right)^2p_{\Lambda}^3\left(\alpha\right),\\
\vdots\quad\vdots\quad\vdots\quad\vdots\quad\vdots\quad\vdots\quad\vdots\quad\vdots\quad\vdots\quad\vdots\quad\vdots\quad\vdots\quad\vdots\quad\vdots\\
p_{\Lambda}\left(\alpha\right)=\alpha \left[p_{\Lambda}^{2\Lambda}\left(\alpha\right)\right]-\left(\frac{b_{1-\Lambda}}{2} \right)^2p_{\Lambda}^{2\Lambda-1}\left(\alpha\right).
\end{array}
\label{recurre1}
\ee
So the claim is true because of (\ref{recurre1}) and the following
\begin{teorema}\emph{The Favard theorem, \cite{Freud} (p. 60)}\label{teoFavard}\\
Let $\{p_n(x)=x^n+\cdots\}$ $(n=0,1,\cdots)$ be a sequence of polynomials with real coefficients, satisfying a recursion formula 
\be
p_{n}(x)=\left(x-\beta_{n}\right)p_{n-1}(x)-\Sigma_np_{n-2}(x)\label{recurr2}
\ee
with positive $\Sigma_n$ and real $\beta_n$; then there exists a distribution $d\alpha$ such that
$$
\int_{-\infty}^{+\infty}{p_n(x)p_m(x)d\alpha(x)}=0 \quad (m\neq n).
$$\label{teo1}
\end{teorema}
\begin{teorema}\emph{\cite{Szego} (p. 44)}\label{teodistinct}\\
The zeros of the orthogonal polynomials $p_n(x)$, associated with distribution $d\alpha(x)$ on the interval $[a,b]$ are distinct and are located in the interior of the interval $[a,b]$.\label{teo2}
\end{teorema}
\subsubsection{Proof of item $(C)$ in theorem \ref{teospec1}}
First of all, we have to recall that $\rho(A)=\|A\|_2$ for every symmetric matrix $A$, where $\rho(A)$ is the spectral radius, i.e.
$$
\rho(A):=max\left\{|\lambda_j| : \lambda_j\in\Sigma_A \right\}.
$$
From $1\leq b_n\leq \sqrt{1+\frac{\Lambda(\Lambda-1)}{k(\Lambda)}}$ and {proposition \ref{propoineqn2}} we can infer
$$
\alpha_1\left(\Lambda\right)=\left\|X^{\Lambda}\right\|_2\leq \sqrt{1+\frac{\Lambda(\Lambda-1)}{k(\Lambda)}}\left\|P_{2\Lambda+1}\left(0,\frac{1}{2},\frac{1}{2} \right)\right\|_2= \sqrt{1+\frac{\Lambda(\Lambda-1)}{k(\Lambda)}}\cos{\left(\frac{\pi}{2\Lambda+2} \right)}
$$
and
$$
\alpha_1\left(\Lambda+1\right)=\left\|X^{\Lambda+1}\right\|_2\geq \left\|P_{2\Lambda+3}\left(0,\frac{1}{2},\frac{1}{2}\right)\right\|_2=\cos{\left(\frac{\pi}{2\Lambda+4}\right)}.
$$
On the other hand, by algebraic calculations, one can easily see that
$$
\sqrt{1+\frac{\Lambda(\Lambda-1)}{k(\Lambda)}}\cos{\left(\frac{\pi}{2\Lambda+2} \right)}\leq \cos{\left(\frac{\pi}{2\Lambda+4}\right)}
$$
is equivalent to
\begin{equation*}
\begin{split}
k\left(\Lambda\right)&\geq \frac{\Lambda(\Lambda-1) \cos^2{\left(\frac{\pi}{2\Lambda+2}\right)}}{\cos^2{\left(\frac{\pi}{2\Lambda+4}\right)}-\cos^2{\left(\frac{\pi}{2\Lambda+2}\right)}}\\
&= \frac{\Lambda(\Lambda-1) \cos^2{\left(\frac{\pi}{2\Lambda+2}\right)}}{
2\sin{\left(\frac{\pi(2\Lambda+3)}{(2\Lambda+2)(2\Lambda+4)} \right)}
\sin{\left(\frac{\pi}{(2\Lambda+2)(2\Lambda+4)} \right)}\left[\cos{\left(\frac{\pi}{2\Lambda+4}\right)}+\cos{\left(\frac{\pi}{2\Lambda+2}\right)}\right]}.
\end{split}
\end{equation*}
And using
$$
\frac{a+1}{a(a+2)}>\frac{1}{1+a},\quad 
\frac{\cos^2{\left(\frac{\pi}{2\Lambda+2}\right)}}{\cos{\left(\frac{\pi}{2\Lambda+4}\right)}+\cos{\left(\frac{\pi}{2\Lambda+2}\right)}}\leq \frac 12\quad\forall\Lambda\in\mathbb{N}\mbox{ and }\sin{x}\geq x^2\quad \forall x\in\left[0,\frac{1}{2}\right],
$$
we obtain
\begin{equation*}
\begin{split}
\frac{\Lambda(\Lambda-1) \cos^2{\left(\frac{\pi}{2\Lambda+2}\right)}}{
2\sin{\left(\frac{\pi(2\Lambda+3)}{(2\Lambda+2)(2\Lambda+4)} \right)}
\sin{\left(\frac{\pi}{(2\Lambda+2)(2\Lambda+4)} \right)} \left[\cos{\left(\frac{\pi}{2\Lambda+4}\right)}+\cos{\left(\frac{\pi}{2\Lambda+2}\right)}\right]}&< \frac{\Lambda(\Lambda-1)}{4\left(\frac{\pi}{2\Lambda+3} \frac{\pi}{(2\Lambda+2)(2\Lambda+4)}\right)^2}\\
&< \frac1{4\pi^4}\Lambda(\Lambda-1) (2\Lambda+2)^2(2\Lambda+3)^2(2\Lambda+4)^2.
\end{split}
\end{equation*}
According to this, 
$$
k(\Lambda)\geq \frac1{4\pi^4}\Lambda(\Lambda-1) (2\Lambda+2)^2(2\Lambda+3)^2(2\Lambda+2)^4\quad\Rightarrow\quad \alpha_1\left(\Lambda\right)< \alpha_1\left(\Lambda+1\right)\quad \forall\Lambda\in\mathbb{N}.
$$
\subsubsection{Proof of item $(D)$ in theorem \ref{teospec1}}\label{proofteodense1}
The scheme of the proof is the following:
\begin{itemize}
\item We firstly prove
\be 
\lim_{\Lambda\rightarrow+\infty}{\alpha_1\left(\Lambda\right)}=1.\label{limmax1}
\ee
\item Then we note that, in the limit $\Lambda\rightarrow+\infty$, $X^{\Lambda}$ can be approximated by $P_{\Lambda}\left(0,\frac{1}{2},\frac{1}{2}\right)$; so we consider the spectra of both matrices.
\item For every $\Lambda\in\mathbb{N}$ we define a continuous, odd and increasing (with respect to $x$) function $G_{\Lambda}(x)$ mapping one spectrum into the other.
\item Through {lemma \ref{lemmadelta}} and {lemma \ref{lemmasigma}} we prove {theorem \ref{teoA01}}, which tells us that $\lim_{\Lambda\rightarrow+\infty} G_{\Lambda}(x)=x$ $\forall x\in[-1,1]$.
\item Finally, in {theorem \ref{convunifGL}}, we prove that $G_{\Lambda}\rightarrow I$ uniformly, and this trivially implies the claim of $(D)$.
\end{itemize}

As for the previous proof, from
$$
\frac{1}{2}\leq \frac{b_n}{2}\leq \frac{\sqrt{1+\frac{\Lambda(\Lambda-1)}{k}}}{2}\quad \forall n\in\left\{\Lambda,\Lambda-1,\cdots,2-\Lambda,1-\Lambda \right\}
$$
and {proposition \ref{propoineqn2}} we obtain
$$
\left\|P_{2\Lambda+1}\left(0,\frac{1}{2},\frac{1}{2} \right) \right\|_2\leq \left\|X^{\Lambda}\right\|_2\leq \left\|P_{2\Lambda+1}\left(0,\frac{\sqrt{1+\frac{\Lambda(\Lambda-1)}{k}}}{2},\frac{\sqrt{1+\frac{\Lambda(\Lambda-1)}{k}}}{2}\right) \right\|_2,
$$
which is equivalent to
\be\label{carabinieri1}
\cos{\left(\frac{\pi}{2\Lambda+2}\right)}\leq \alpha_1\left(\Lambda\right)\leq \sqrt{1+\frac{\Lambda(\Lambda-1)}{k}}\cos{\left(\frac{\pi}{2\Lambda+2}\right)},
\ee
this and $k=k\left(\Lambda\right)\geq \Lambda^2\left(\Lambda+1\right)^2$ concludes the proof of (\ref{limmax1}).

The inequality (\ref{limitvaluesalpha1})$_2$ follows trivially from (\ref{carabinieri1}), $\cos{x}\geq 1-\frac{x^2}{2}$ $\forall x\in[0,1]$ and $\frac{\pi}{2\Lambda+2}\leq 1$ $\forall\Lambda\in\mathbb{N}.$

{Corollary 6.3.8} in \cite{Horn} p. 370 states that (here $M_n$ is the space of $n\times n$ complex matrices)

\emph{
Let $A,E\in M_n$, assume that $A$ is Hermitian and that $A+E$ is normal, let $\{\lambda_1,\cdots,\lambda_n\}$ be the eigenvalues of $A$ arranged in increasing order $\left(\lambda_1\leq \lambda_2\leq \cdots\leq \lambda_n\right)$ and let $\left\{\widehat{\lambda}_1,\cdots,\widehat{\lambda}_n\right\}$ be the eigenvalues of $A+E$, ordered so that $Re\left( \widehat{\lambda}_1\right)\leq Re\left(\widehat{\lambda}_2\right)\leq \cdots\leq Re\left(\widehat{\lambda}_n\right)$. Then}
\be\label{ineqAE}
\left[\sum_{i=1}^{n}\left\vert \widehat{\lambda}_i-\lambda_i\right\vert^2 \right]^{\frac{1}{2}}\leq \left\|E\right\|_2.
\ee
According to this, setting $A:=P_{2\Lambda+1}\left(0,\frac{1}{2},\frac{1}{2}\right)$, $E:=X^{\Lambda}-P_{2\Lambda+1}\left(0,\frac{1}{2},\frac{1}{2}\right)$, then $A$ and $A+E$ are both symmetric, so (\ref{ineqAE}) becomes
$$
\left[\sum_{i=1}^{2\Lambda+1}\left\vert \alpha_i(\Lambda)-\widetilde{\alpha}_i(\Lambda)\right\vert^2 \right]^{\frac{1}{2}}\leq \left\|E\right\|_2.
$$
From $\sqrt{1+x}\leq 1+\frac{x}{2}$, $k=k\left(\Lambda\right)\geq \Lambda^2\left(\Lambda+1\right)^2$ and $|n|\leq \Lambda$ we obtain
$$
\frac{1}{2}\left[\sqrt{1+\frac{n(n-1)}{k}}-1\right]\leq \frac{n(n-1)}{4k}\leq \frac{1}{4(\Lambda+1)^2},
$$
so {proposition \ref{propoineqn2}} implies
$$
\left\|E\right\|_2\leq \left\| P_{2\Lambda+1}\left(0,\frac{1}{4(\Lambda+1)^2},\frac{1}{4(\Lambda+1)^2}\right)\right\|_2=\frac{1}{2(\Lambda+1)^2}\cos{\left(\frac{\pi}{2\Lambda+2} \right)}< \frac{1}{2(\Lambda+1)^2}
$$
and then
\be\label{boundsumalpha1}
\left[\sum_{i=1}^{2\Lambda+1}\left\vert \alpha_i(\Lambda)-\widetilde{\alpha}_i(\Lambda)\right\vert^2 \right]^{\frac{1}{2}}< \frac{1}{2(\Lambda+1)^2}\quad \forall\Lambda.
\ee

For every $\Lambda\in\mathbb{N}$ we can define a continuous function $G_{\Lambda}:[-1,1]\rightarrow [-\alpha_1\left(\Lambda\right),\alpha_1\left(\Lambda\right)]$ such that $G_{\Lambda}\left[\widetilde{\alpha}_n\left(\Lambda\right) \right]=\alpha_n\left(\Lambda\right)$, $G_{\Lambda}(-x)=-G_{\Lambda}(x)$, $G_{\Lambda}(x)=\alpha_1\left(\Lambda\right)$ $\forall x\in\left[\widetilde{\alpha}_1\left(\Lambda\right),1\right]$, for instance we can join two ``consecutive'' points $\left(\widetilde{\alpha}_i\left(\Lambda\right),\alpha_i\left(\Lambda\right) \right)$ and $\left(\widetilde{\alpha}_{i+1}\left(\Lambda\right),\alpha_{i+1}\left(\Lambda\right) \right)$ by a straight line; furthermore, because of 
$$
G_{\Lambda}\left[\widetilde{\alpha}_n\left(\Lambda\right) \right]=\alpha_n\left(\Lambda\right)<G_{\Lambda}\left[\widetilde{\alpha}_{n-1}\left(\Lambda\right) \right]=\alpha_{n-1}\left(\Lambda\right),
$$
we can assume that every function $G_{\Lambda}(x)$ is also increasing with respect to $x$.

The $G_{\Lambda}(x)$ are all odd functions so we can restrict our attention to the $x\in[0,1]$, but it is also true that the continuity and the monotonicity of every $G_{\Lambda}$ imply that 
$$
\forall \varepsilon>0, \forall x\in[0,1] \exists \delta=\delta(\varepsilon,\Lambda,x) \mbox{ s.t. } y\in[0,1],\begin{cases}
|x-y|<\delta \Rightarrow \left\vert G_{\Lambda}(x)-G_{\Lambda}(y)\right\vert <\varepsilon,\\
|x-y|>\delta \Rightarrow \left\vert G_{\Lambda}(x)-G_{\Lambda}(y)\right\vert >\varepsilon.
\end{cases}
$$
At this point we need to prove the following
\begin{lemma}\label{lemmadelta}
Let $\varepsilon>0$ and $\overline{x}\in[0,1]$ such that
$$
{\limsup}_{\Lambda\rightarrow +\infty}\left\vert \overline{x}-G_{\Lambda}\left(\overline{x}\right) \right\vert=0,
$$
then
\be
{\liminf}_{\Lambda\rightarrow +\infty}\delta(\varepsilon,\Lambda,\overline{x})=\widetilde{\delta}(\varepsilon,\overline{x})>0.
\ee
\proof
Let $\varepsilon>0$ and assume, per absurdum, that
$$
{\liminf}_{\Lambda\rightarrow +\infty}\delta(\varepsilon,\Lambda,\overline{x})=0,
$$
then we can find a sequence $\left\{\widetilde{\Lambda}_n\right\}_{n\in\mathbb{N}}$ such that
\be\label{lim_delta_0}
\lim_n \delta(\varepsilon, \widetilde{\Lambda}_n,\overline{x})=0
\ee
and, correspondingly, because of (\ref{lim_delta_0}) we can assume that $n$ is sufficiently large so that we can find $x\in[0,1]$ with $\frac{\varepsilon}{4}>\vert \overline{x}-x\vert > \delta(\varepsilon, \widetilde{\Lambda}_n,\overline{x})$, $\vert \overline{x}-G_{\widetilde{\Lambda}_n}(\overline{x})\vert<\frac{\varepsilon}{4}$  and $\vert G_{\widetilde{\Lambda}_n}(\overline{x})-G_{\widetilde{\Lambda}_n}(x)\vert >\varepsilon$; then
\begin{equation*}
\begin{split}
\vert x-G_{\widetilde{\Lambda}_n}(x) \vert &=\vert x-\overline{x}+\overline{x}-G_{\widetilde{\Lambda}_n} (\overline{x})+ G_{\widetilde{\Lambda}_n}(\overline{x})-G_{\widetilde{\Lambda}_n}(x) \vert\\
& \geq  \vert G_{\widetilde{\Lambda}_n}(\overline{x})-G_{\widetilde{\Lambda}_n}(x) \vert -\vert \overline{x}-x\vert -\vert \overline{x}-G_{\widetilde{\Lambda}_n}(\overline{x})\vert\\
& \geq \varepsilon- \frac{\varepsilon}{2}=\frac{\varepsilon}{2}.
\end{split}
\end{equation*}
This last inequality and (\ref{valuecos}) imply that there exist a finite set of indices $I$ with $|I|=m(n)$ such that the correspondings eigenvalues of $P_{2\widetilde{\Lambda}_n+1}\left(0,\frac{1}{2},\frac{1}{2}\right)$, in symbols $\left\{\widetilde{\alpha}_i\left(\widetilde{\Lambda}_n \right)\right\}_{i\in I}$, fulfill
$$
\frac{\varepsilon}{4}>\left\vert \overline{x}-\widetilde{\alpha}_i \left(\widetilde{\Lambda}_n \right)\right\vert > \delta(\varepsilon, \widetilde{\Lambda}_n,\overline{x})\quad \forall i\in I  \Longrightarrow \left\vert \widetilde{\alpha}_i \left(\widetilde{\Lambda}_n \right)-G_{\widetilde{\Lambda}_n}\left[\widetilde{\alpha}_i \left(\widetilde{\Lambda}_n \right)\right] \right\vert>\frac{\varepsilon}{2}\quad \forall i\in I
$$
and of course (\ref{lim_delta_0}) implies that $m(n)\overset{n\rightarrow +\infty}\longrightarrow +\infty$, so
$$
\lim_n\left[\sum_{i\in I}\left\vert \widetilde{\alpha}_i \left(\widetilde{\Lambda}_n \right)-G_{\widetilde{\Lambda}_n}\left[\widetilde{\alpha}_i \left(\widetilde{\Lambda}_n \right) \right] \right\vert^2\right]=+\infty,
$$
which disagrees with (\ref{boundsumalpha1}), so the proof is finished.
\endproof
\end{lemma}

Let
$$
A:=\left\{x\in[0,1]:{\limsup}_{\Lambda\rightarrow +\infty}\left\vert x-G_{\Lambda}(x) \right\vert=0 \right\}
$$
we have that $0\in A$ and also that $1\in A$ because
$$
\lim_{\Lambda\rightarrow+\infty}\alpha_{1}\left(\Lambda\right)=\lim_{\Lambda\rightarrow+\infty}\widetilde{\alpha}_{1}\left(\Lambda\right)=\lim_{\Lambda\rightarrow+\infty}G_{\Lambda}\left[\widetilde{\alpha}_{1}\left(\Lambda\right)\right]=1.
$$
In order to prove item $(D)$ in {theorem \ref{teospec1}} we need the following
\begin{lemma}\label{lemmasigma}
If $0\leq\overline{x}\leq 1,\overline{x}\in A$, then $\exists\sigma>0$ such that $x\in\left]\max\left\{\overline{x}-\sigma,0\right\},\min\left\{\overline{x}+\sigma,1\right\}\right[\Longrightarrow x\in A$.
\proof
Let $\varepsilon>0$, then {lemma \ref{lemmadelta}} implies
$$
\liminf_{\Lambda\rightarrow+\infty}\delta\left(\varepsilon,\Lambda,\overline{x}\right)=\widetilde{\delta}\left(\varepsilon,\overline{x}\right)>0;
$$
so, if we set $\sigma:=\min\left\{\frac{\delta\left(\varepsilon,\overline{x}\right)}{2},\varepsilon\right\}$ and we take $x\in\left]\max\left\{\overline{x}-\sigma,0\right\},\min\left\{\overline{x}+\sigma,1\right\}\right[$, then
\begin{equation*}
\begin{split}
\limsup_{\Lambda\rightarrow+\infty}\left\vert x-G_{\Lambda}(x) \right\vert&=\limsup_{\Lambda\rightarrow+\infty}\left\vert x-G_{\Lambda}(x) -\overline{x}+\overline{x}-G_{\Lambda}\left(\overline{x}\right)+G_{\Lambda}\left(\overline{x}\right) \right\vert\\
&\leq \limsup_{\Lambda\rightarrow+\infty}\left\vert x-\overline{x}\right\vert + \left\vert \overline{x}-G_{\Lambda}\left(\overline{x}\right)\right\vert+\left\vert G_{\Lambda}(x)-G_{\Lambda}\left(\overline{x}\right)\right\vert\leq 2\varepsilon,
\end{split}
\end{equation*}
of course $\varepsilon$ can be chosen arbitrary small, so the proof is finished.
\endproof
\end{lemma}
According to this, we can trivially infer that
\begin{corollario}\label{corollAB}
$$
A=[0,1]
$$
or
$$
A=[0,x_1[\cupdot ]x_2,x_3[\cupdot \cdots \cupdot ]x_s,1]\quad\mbox{and}\quad B:=[0,1]\setminus A=[x_1,x_2]\cupdot [x_3,x_4]\cupdot \cdots,
$$
where $x_1<x_2<x_3<x_4\cdots $ are suitable points of $]0,1[$.
\end{corollario}
We are now ready to prove the following
\begin{teorema}\label{teoA01}
$$
A=[0,1]
$$
\proof
Let us assume, per absurdum, that $A\neq [0,1]$, then {corollary \ref{corollAB}} implies
\be\label{decompAsigma}
B:=[0,1]\setminus A=[x_1,x_2]\cupdot [x_3,x_4]\cupdot \cdots,
\ee
so if $x\in A$, $\delta>0$, $x_1-\delta<x<x_1$ and $\limsup_{\Lambda\rightarrow+\infty}\left\vert x_1-G_{\Lambda}\left(x_1\right)\right\vert=k>0$, then
\be
\begin{split}
\limsup_{\Lambda\rightarrow+\infty}\left\vert G_{\Lambda}\left(x\right)-G_{\Lambda}\left(x_1\right)\right\vert&= \limsup_{\Lambda\rightarrow+\infty}\left\vert G_{\Lambda}\left(x\right)-x_1+x_1-G_{\Lambda}\left(x_1\right)\right\vert \\
&\leq  \limsup_{\Lambda\rightarrow+\infty}\left\vert G_{\Lambda}\left(x\right)-x_1\right\vert+\left\vert x_1-G_{\Lambda}\left(x_1\right)\right\vert\\
&\leq \delta+k,
\end{split}
\ee
because $x\in A$.

On the other hand
\be
\begin{split}
\limsup_{\Lambda\rightarrow+\infty}\left\vert G_{\Lambda}\left(x\right)-G_{\Lambda}\left(x_1\right)\right\vert&= \limsup_{\Lambda\rightarrow+\infty}\left\vert G_{\Lambda}\left(x\right)-x_1+x_1-G_{\Lambda}\left(x_1\right)\right\vert \\
&\geq  \limsup_{\Lambda\rightarrow+\infty}\left\vert x_1-G_{\Lambda}\left(x_1\right)\right\vert-\left\vert G_{\Lambda}\left(x\right)-x_1\right\vert\\
&\geq k-\delta.
\end{split}
\ee
According to this, we obtain
$$
\lim_{\delta\rightarrow 0}\left[\limsup_{\Lambda\rightarrow+\infty}\left\vert G_{\Lambda}\left(x\right)-G_{\Lambda}\left(x_1\right)\right\vert-k\right]=0
$$
so we can infer that $\limsup_{\Lambda\rightarrow+\infty}G_{\Lambda}\left(x_1\right)=k+x_1$ and we can find a sequence $\left\{\widetilde{\Lambda}_n\right\}_{n\in\mathbb{N}}$ such that
$$
\lim_{n\rightarrow+\infty}G_{\widetilde{\Lambda}_n}\left(x_1\right)=k+x_1, 
$$
but we also know that $G_{\Lambda}(x)$ is increasing with respect to $x$, so
$$
\liminf_{n\rightarrow+\infty}G_{\widetilde{\Lambda}_n}\left(x\right)\geq k+x_1\quad  \forall x\in\left[x_1, x_1+\frac{k}{2}\right].
$$
This implies
$$
\liminf_{n\rightarrow+\infty}\left\vert x-G_{\widetilde{\Lambda}_n}(x)\right\vert\geq k+x_1-\left(x_1+\frac{k}{2}\right)=\frac{k}{2}\quad \forall x\in\left[x_1, x_1+\frac{k}{2}\right].
$$
This last inequality and (\ref{valuecos}) imply that there exist a finite set of indices $I$ with $|I|=m(n)$ such that the correspondings eigenvalues of $P_{2\widetilde{\Lambda}_n+1}\left(0,\frac{1}{2},\frac{1}{2}\right)$, in symbols $\left\{\widetilde{\alpha}_i\left(\widetilde{\Lambda}_n \right)\right\}_{i\in I}$, fulfill
$$
\widetilde{\alpha}_i\left(\widetilde{\Lambda}_n\right)\in\left[x_1, x_1+\frac{k}{2}\right]\quad \forall i\in I \quad \Longrightarrow \left\vert \widetilde{\alpha}_i\left(\widetilde{\Lambda}_n\right)-G_{\widetilde{\Lambda}_n}\left[\widetilde{\alpha}_i\left(\widetilde{\Lambda}_n \right)\right] \right\vert>\frac{k}{4}\quad \forall i\in I
$$
and of course $m(n)\overset{n\rightarrow +\infty}\longrightarrow +\infty$, so
$$
\lim_{n\rightarrow+\infty}\left[\sum_{i\in I}\left\vert \widetilde{\alpha}_i\left(\widetilde{\Lambda}_n\right)-G_{\widetilde{\Lambda}_n}\left[\widetilde{\alpha}_i\left(\widetilde{\Lambda}_n \right)\right] \right\vert^2\right]=+\infty,
$$
which disagrees with (\ref{boundsumalpha1}), so the proof is finished.
\endproof
\end{teorema}
According to this, we have that
$$
\lim_{\Lambda\rightarrow+\infty}G_{\Lambda}(x)=x\quad \forall x\in [0,1],
$$
in the next theorem we will always denote the sequence $\left\{\Lambda\right\}_{\Lambda\in \mathbb{N}}$ and its subsequences with the same notation.
\begin{teorema}\label{convunifGL}
$$
\limsup_{\Lambda\rightarrow +\infty}\left[\max_{x\in [0,1]}\left\{\left\vert x- G_{\Lambda}(x)\right\vert \right\} \right]=0.
$$
\proof
Let us assume, per absurdum, that 
$$
\limsup_{\Lambda\rightarrow +\infty}\left[\max_{x\in [0,1]}\left\{\left\vert x- G_{\Lambda}(x)\right\vert \right\} \right]=M>0,
$$
and set
$$
x_{\Lambda}:=\max_{x\in [0,1]}\left\{\left\vert x- G_{\Lambda}(x)\right\vert \right\};
$$
we have that (up to a suitable subsequence)
$$
\lim_{\Lambda\rightarrow+\infty} \left\vert x_{\Lambda}-G_{\Lambda}\left( x_{\Lambda} \right)\right\vert =M.
$$
The sequence $\left\{ x_{\Lambda}\right\}_{\Lambda\in\mathbb{N}}$ is bounded, so we have that (up to a further suitable subsequence)
$$
\lim_{\Lambda\rightarrow+\infty}x_{\Lambda}=\overline{x}\in [0,1]=A,
$$
at this point, let us choose $\varepsilon$, $x$ so that
$$
0<\varepsilon<\frac{M}{8}\quad,\quad\sigma:=\min\left\{\frac{\widetilde{\delta}\left(\varepsilon,\overline{x}\right)}{2},\frac{M}{8}\right\}>0\quad,\quad x\in\left[\overline{x}-\frac{\sigma}{2},\overline{x}+\frac{\sigma}{2}\right]
$$
and $\Lambda$ such that
$$
\left\vert x_{\Lambda}-G_{\Lambda}\left(x_{\Lambda}\right) \right\vert>\frac{M}{2}\quad,\quad \left\vert x-x_{\Lambda} \right\vert<\sigma \quad,\quad \left\vert \overline{x}-x_{\Lambda} \right\vert<\sigma,
$$
then we obtain (if $\Lambda$ is sufficiently large)
\begin{equation*}
\begin{split}
\left\vert x-G_{\Lambda}\left(x\right)\right\vert &\geq \left\vert x_{\Lambda}-G_{\Lambda}\left(x_{\Lambda}\right) \right\vert-\left\vert x-x_{\Lambda} \right\vert- \left\vert G_{\Lambda}\left(\overline{x} \right)-G_{\Lambda}\left(x_{\Lambda}\right)\right\vert-\left\vert G_{\Lambda}\left(\overline{x} \right)-G_{\Lambda}\left(x\right)\right\vert\\
&>\frac{M}{2}-\sigma-\varepsilon-\varepsilon>\frac{M}{8}.
\end{split}
\end{equation*}
This last inequality implies that there exist a finite set of indices $I$ with $|I|=m(\Lambda)$ such that the correspondings eigenvalues of $P_{2\widetilde{\Lambda}_n+1}\left(0,\frac{1}{2},\frac{1}{2}\right)$, in symbols $\left\{\widetilde{\alpha}_i\left(\Lambda\right) \right\}_{i\in I}$, fulfill
$$
\widetilde{\alpha}_i\left(\Lambda\right)\in\left[\overline{x}-\frac{\sigma}{2},\overline{x}+\frac{\sigma}{2}\right]\quad \forall i\in I \quad \Longrightarrow \left\vert \widetilde{\alpha}_i\left(\Lambda\right)-G_{\Lambda}\left[\widetilde{\alpha}_i\left(\Lambda\right) \right] \right\vert>\frac{M}{8}\quad \forall i\in I
$$
and of course $m(\Lambda)\overset{\Lambda\rightarrow +\infty}\longrightarrow +\infty$, so
$$
\lim_{\Lambda\rightarrow +\infty}\left[\sum_{i\in I}\left\vert \widetilde{\alpha}_i\left(\Lambda\right)-G_{\Lambda}\left[\widetilde{\alpha}_i\left(\Lambda\right)\right] \right\vert^2\right]=+\infty,
$$
which disagrees with (\ref{boundsumalpha1}), so the proof is finished.
\endproof
\end{teorema}
We are now ready to complete the proof of item $(D)$ of {theorem \ref{teospec1}}, because if  $\varepsilon>0$ the last theorem implies that there exists a $\widetilde{\Lambda}=\widetilde{\Lambda}\left(\varepsilon\right)$ such that $\left\vert x-G_{\Lambda}\left(x\right)\right\vert<\varepsilon$ $\forall \Lambda>\widetilde{\Lambda}$ and $\forall x\in [0,1]$, while (\ref{valuecos}) implies 
\begin{equation*}
\begin{split}
\left\vert \widetilde{\alpha}_{n+1}\left(\Lambda\right)-\widetilde{\alpha}_n\left(\Lambda\right)\right\vert&=\left\vert \cos{\left[\frac{(n+1)\pi}{\Lambda+1} \right]}-\cos{\left(\frac{n\pi}{\Lambda+1} \right)}  \right\vert\\
&=
\left\vert
2 \sin{\left[\frac{(2n+1)\pi}{\Lambda+1} \right]} \sin{\left(\frac{\pi}{\Lambda+1} \right)}
\right\vert\leq 2 \sin{\left(\frac{\pi}{\Lambda+1} \right)},
\end{split}
\end{equation*}
this means that there exists a $\widehat{\Lambda}=\widehat{\Lambda}\left(\varepsilon\right)$ such that $\left\vert \widetilde{\alpha}_i\left(\Lambda\right)-\widetilde{\alpha}_{i+1}\left(\Lambda\right)\right\vert<\varepsilon$ $\forall \Lambda>\widetilde{\Lambda}$, $\forall i$. 

Finally, if we set $\overline{\Lambda}\left(\varepsilon\right)=\max\left\{\widehat{\Lambda}\left(\varepsilon\right),\widetilde{\Lambda}\left(\varepsilon\right) \right\}$, then $\forall \Lambda>\overline{\Lambda}$ we obtain
\begin{equation*}
\begin{split}
\left\vert \alpha_{i}\left(\Lambda\right)-\alpha_{i+1}\left(\Lambda\right) \right\vert &\leq \left\vert \alpha_{i}\left(\Lambda\right)-\widetilde{\alpha}_{i}\left(\Lambda\right) \right\vert+\left\vert \alpha_{i+1}\left(\Lambda\right)-\widetilde{\alpha}_{i+1}\left(\Lambda\right) \right\vert+\left\vert \widetilde{\alpha}_{i}-\widetilde{\alpha}_{i+1}\right\vert\\
&=\left\vert G_{\Lambda}\left[\widetilde{\alpha}_{i}\left(\Lambda\right)\right]-\widetilde{\alpha}_{i}\left(\Lambda\right) \right\vert+\left\vert G_{\Lambda}\left[ \widetilde{\alpha}_{i+1}\left(\Lambda\right)\right]-\widetilde{\alpha}_{i+1}\left(\Lambda\right) \right\vert+\left\vert \widetilde{\alpha}_{i}-\widetilde{\alpha}_{i+1}\right\vert\\
&<\varepsilon+\varepsilon+\varepsilon=3\varepsilon,
\end{split}
\end{equation*}
so the proof is completed.

\subsection{The proofs of theorems of section \ref{spherespec}}

\subsubsection{Proof of item $(A)$ in theorem \ref{teospec2}}

Consider the unitary and involutive operator $U_0=U_0^\dagger=U_0^{-1}$
corresponding to the inversion operator of the $x_3$-axis (this exists
by the $O(3)$-covariance of our model): $U_0\,x_0\,U_0=-x_0$,
$U_0x_\pm U_0=x_\pm$. Then $x_0\bm{\chi}=\alpha \bm{\chi}$ implies
 $x_0(U_0\bm{\chi})=-\alpha (U_0\bm{\chi})$, i.e. $U_0\bm{\chi}$ is
an eigenvector of $x_0$ with the opposite eigenvalue.
\subsubsection{Proof of item $(B)$ in theorem \ref{teospec2}}
According to the last proof, we can equivalently set $M_{m}\left(\Lambda;\alpha\right):=B_m\left(\Lambda\right)+\alpha I_{\Lambda-m+1}$, then the eigenvalue problem for $B_m\left(\Lambda\right)$ is equivalent to solve $\det{\left[M_m(\Lambda;\alpha)\right]}=0$; in order to do this we define $M_m^h$ as the $h\times h$ submatrix of $M_{m}$ formed by the first $h$ rows and columns, then
$$p_{n(\Lambda;m)}(\alpha):=\det{\left[M_{m}(\Lambda;\alpha)\right]}\quad\mbox { and }\quad p_{n(\Lambda;m_1)}^{h}(\alpha):=\det\left\{M_{m_1}^{h}\left(\Lambda;\alpha\right)\right\},$$
where $n(\Lambda;m'):=\Lambda-|m'|+1$ is the degree of the polynomial $p_{n(\Lambda,m')}$.

It is not difficult to see that
\begin{itemize}
\item when $n=1\Leftrightarrow |m|=\Lambda$, then $\alpha=0$;
\item when $n=2$, then
$$
\left|
\begin{array}{cc}
\alpha&c_{\Lambda}A_{\Lambda}^{0,\Lambda-1}\\
c_{\Lambda}A_{\Lambda}^{0,\Lambda-1}&\alpha\\
\end{array}
\right|=\alpha^2-\left(c_{\Lambda}A_{\Lambda}^{0,\Lambda-1}\right)^2=:p_2\left(\alpha\right)\Rightarrow \alpha_{1,2}=\pm c_{\Lambda}A_{\Lambda}^{0,\Lambda-1};
$$
\item when $n=3$, then
$$
\left|
\begin{array}{ccc}
\alpha&c_{\Lambda-1}A_{\Lambda-1}^{0,\Lambda-2}&0\\
c_{\Lambda-1}A_{\Lambda-1}^{0,\Lambda-2}&\alpha&c_{\Lambda}A_{\Lambda}^{0,\Lambda-2}\\
0&c_{\Lambda}A_{\Lambda}^{0,\Lambda-2}&\alpha\\
\end{array}
\right|=\alpha\left[\alpha^2-\left(c_{\Lambda}A_{\Lambda}^{0,\Lambda-2}\right)^2\right]-\alpha\left( c_{\Lambda-1}A_{\Lambda-1}^{0,\Lambda-2}\right)^2$$
$$
=:p_3\left(\alpha\right);
$$
\item in general, let $\widetilde{n}=n\left(\Lambda;\widetilde{m}\right)$, then one can calculate $p_{\widetilde{n}}\left(\alpha\right)$ through the use of this recursion formula 
\end{itemize}
\be
\begin{array}{c}
p_{\widetilde{n}}^{2}\left(\alpha\right):=\det{\left\{M_{\widetilde{m}}^2\left(\Lambda;\alpha\right)\right\}},\\
p_{\widetilde{n}}^{3}\left(\alpha\right):=\det{\left\{M_{\widetilde{m}}^3\left(\Lambda;\alpha\right)\right\}},\\
p_{\widetilde{n}}^{4}\left(\alpha\right):=\alpha \left[p_{\widetilde{n}}^{3}\left(\alpha\right)\right]-\left(c_{\widetilde{m}+3}A_{\widetilde{m}+3}^{0,\widetilde{m}} \right)^2p_{\widetilde{n}}^{2}\left(\alpha\right),\\
p_{\widetilde{n}}^{5}\left(\alpha\right):=\alpha \left[p_{\widetilde{n}}^{4}\left(\alpha\right)\right]-\left(c_{\widetilde{m}+4}A_{\widetilde{m}+4}^{0,\widetilde{m}} \right)^2p_{\widetilde{n}}^{3}\left(\alpha\right),\\
\vdots\quad\vdots\quad\vdots\quad\vdots\quad\vdots\quad\vdots\quad\vdots\quad\vdots\quad\vdots\quad\vdots\quad\vdots\quad\vdots\quad\vdots\quad\vdots\quad\vdots\quad\vdots\\
p_{\widetilde{n}}\left(\alpha\right)=\alpha \left[p_{\widetilde{n}}^{\widetilde{n}-1}\left(\alpha\right)\right]-\left(c_{\Lambda}A_{\Lambda}^{0,\widetilde{m}} \right)^2p_{\widetilde{n}}^{\widetilde{n}-2}\left(\alpha\right).
\end{array}
\label{recurre}
\ee
Then the proof of item $(B)$ follows trivially from (\ref{recurre}), {theorem \ref{teoFavard}} and {theorem \ref{teodistinct}}, as for {section \ref{proofteosimple1}}.
\subsubsection{Proof of (\ref{disalpha}) in theorem \ref{teospec2}}
In this proof we will use the following theorem (here $\left\{p_n\right\}_{n\in\mathbb{N}}$ is a sequence of orthogonal polynomials):
\begin{teorema}\emph{\cite{Szego} (p. 46)}\label{teointerlace}\\
Let $x_1<x_2<\cdots<x_2$ be the zeros of $p_n(x)$. Then each interval $[x_{\nu},x_{\nu+1}]$ contains exactly one zero of $p_{n+1}(x)$.\label{teo3}
\end{teorema}
Anyway, this is the scheme of the proof:
\begin{itemize}
\item We firstly use {theorem \ref{teo1}} to prove that there exist a $\mathbb{R}$-measure such that the polynomials $\left\{p_{n(\Lambda;m)}^{h}\right\}_{h=1}^{n(\Lambda;m)}$ are orthogonal with respect to that measure; this implies that we can apply {theorem \ref{teo2}} obtaining that all the roots of every polynomial $p_{n(\Lambda;m)}^{h}$ are real and simple.
\item Then we use {lemma \ref{lemmadisugA}} and {theorem \ref{teointerlace}} to prove also that 
$$\rho\left(B_m\right)=\|B_{m}\|_2<\|B_{m-1}^{n\left(\Lambda;m\right)}\|_2=\rho\left(B_{m-1}^{n\left(\Lambda;m\right)} \right),$$
where $\rho$ is the spectral radius.
\item This last inequality involving the spectral radii trivially implies (\ref{disalpha}).
\end{itemize}
According to this, let's start with the first point of this scheme.
\begin{lemma}
The roots of $p_{n(\Lambda;m)}^h$ are real and simple, and if $\alpha^{\nu}_1(\Lambda;m)>\alpha^{\nu}_2(\Lambda;m)>\cdots>\alpha^{\nu}_{\nu}(\Lambda;m)$ are the zeros of $p_{n(\Lambda;m)}^{\nu}(\alpha)$, then every interval $\left[\alpha^{\nu+1}_{i+1}(\Lambda;m), \alpha^{\nu+1}_{i}(\Lambda;m)\right]$ contains exactly one zero of $p_{n(\Lambda;m)}^{\nu}(\alpha)$.
\label{lemma1}
\proof
The matrices $B_{m}^h\left(\Lambda\right)$ are all symmetric, so the roots of $p_{n(\Lambda;m)}^{h}(\alpha)$ are real; while the sequence of polynomials $\left\{p_{n(\Lambda;m)}^h\right\}_{h=1}^{n(\Lambda;m)}$ fulfill the recurrence relation (\ref{recurre}) and because of {theorem \ref{teo1}} we can infer that there exists a distribution $d\Theta(\alpha)$ such that
$$
\int_{-\infty}^{+\infty}{p_{n(\Lambda;m)}^j(\alpha)p_{n(\Lambda;m)}^h(\alpha)d\Theta(\alpha)}=0 \quad (j\neq h).
$$
Finally, we can apply {theorem \ref{teo2}} and {theorem \ref{teo3}} to the set $\{p_{n(\Lambda;m)}^h(\alpha)\}_{h=1}^{n(\Lambda;m)}$ of polynomials, so the proof is finished.
\endproof
\end{lemma}
We firstly prove an inequality involving the $B_m$-matrix elements, which implies the aforementioned inequality between the spectral radii.
\begin{lemma}
Let 
\be 
1\leq m\leq \Lambda,\quad j\in\mathbb{N}_0, \quad 1\leq l:=m+j\leq\Lambda;\label{condidisugA}
\ee
then
\be 
c_{l}A_{l}^{0,m-1}>c_{l+1}A_{l+1}^{0,m}.\label{disugA}
\ee 
\label{lemmadisugA}
\proof
Because of (\ref{Clebsch}) and (\ref{defcl}), we obtain
$$
c_{l}A_{l}^{0,m-1}=\sqrt{1+\frac{l^2}{k}}\sqrt{\frac{(l+m-1)(l-m+1)}{4l^2-1}}
$$
and
$$
c_{l+1}A_{l+1}^{0,m}=\sqrt{1+\frac{(l+1)^2}{k}}\sqrt{\frac{(l+m+1)(l-m+1)}{4(l+1)^2-1}},
$$
then (\ref{disugA}) becomes
$$
\left(1+\frac{l^2}{k}\right)\left(\frac{l+m-1}{4l^2-1}\right)-\left(1+\frac{(l+1)^2}{k}\right)\left(\frac{l+m+1}{4(l+1)^2-1}\right)>0
$$
$\forall$ $1\leq m\leq\Lambda$ and $1\leq l \leq \Lambda $; by algebraic calculations, one can prove that the last inequality is equivalent to the following one:
\be
\underbrace{\left[k+l^2\right](l+m-1)(2l+3)}_{A}-\underbrace{\left[k+(l+1)^2\right](l+m+1)(2l-1)}_{B}>0
\label{disequivalente}
\ee
$\forall$ $1\leq m\leq\Lambda$ and $1\leq l \leq \Lambda $.

Furthermore, one has
\bea
A=&&{\color{red}{2kl^2}}+{\color{blue}{2klm}}+{\color{green}{kl}}+{\color{orange}{3km}}-{\color{purple}{3k}}+{\color{brown}{2l^4}}+{\color{gray}{2l^3m}}+{\color{cyan}{l^3}}+{\color{drab}{3l^2m}}-{\color{olive}3l^2},\nn
B=&&{\color{red}{2kl^2}}+{\color{blue}{2klm}}+{\color{green}{kl}}-{\color{orange}{km}}-{\color{purple}{k}}+{\color{brown}{2l^4}}+{\color{gray}{2l^3m}}+{\color{cyan}5l^3}+{\color{drab}3l^2m}+{\color{olive}3l^2}-{\color{violet}l}-{\color{teal}m}-{\color{darkgray}1};\nonumber
\eea
finally, (\ref{disequivalente}) becomes
$$
A-B={\color{orange}{4km}}-{\color{purple}{2k}}-{\color{cyan}{4l^3}}-{\color{olive}6l^2}+{\color{violet}l}+{\color{teal}m}+{\color{darkgray}1}>0
$$
$\forall$ $1\leq m\leq\Lambda$ and $1\leq l \leq \Lambda $.

From $k\left(\Lambda\right)\geq \Lambda^2\left(\Lambda+1\right)^2$ we obtain
$$
4km-2k-4l^3-6l^2+l+m+1\geq 2\Lambda^2(\Lambda+1)^2-4\Lambda^3-6\Lambda^2=2\Lambda^2\left(\Lambda^2-2\right)> 0\quad\forall\Lambda\geq 2,
$$
while when $\Lambda=1$
$$
4km-2k-4l^3-6l^2+l+m+1\geq 2[1^2(2)^2]-4-6+3=1,
$$
so the proof is finished.
\endproof
\end{lemma}
\begin{lemma}
Let $m\geq 1$, then
$$
\|B_{m}\|_2<\|B_{m-1}^{n\left(\Lambda;m\right)}\|_2.
$$
\label{lemmadisugB}
\proof
The matrices $B_{m}$ and $B_{m-1}^{n\left(\Lambda;m\right)}$ have the same dimensions, they are explicitly
\be\label{matr1}
B_{m}=
\left(
\begin{array}{ccccccc}
0&c_{m+1}A_{m+1}^{0,m}&0&0&\cdots&0&0\\
c_{m+1}A_{m+1}^{0,m}&0&c_{m+2}A_{m+2}^{0,m}&0&\cdots&0&0\\
0&c_{m+2}A_{m+1}^{0,m}&0&c_{m+3}A_{m+3}^{0,m}&\cdots&0&0\\
\vdots&\vdots&\vdots&\vdots&\ddots&\vdots&\vdots\\
0&0&0&0&\cdots&0&c_{\Lambda}A_{\Lambda}^{0,m}\\
0&0&0&0&\cdots&c_{\Lambda}A_{\Lambda}^{0,m}&0\\
\end{array}
\right)
\ee
and 
\be\label{matr2}
B_{m-1}^{n\left(\Lambda;m\right)}=
\left(
\begin{array}{ccccccc}
0&c_{m}A_{m}^{0,m-1}&0&0&\cdots&0&0\\
c_{m}A_{m}^{0,m-1}&0&c_{m+1}A_{m+1}^{0,m-1}&0&\cdots&0&0\\
0&c_{m+1}A_{m+1}^{0,m-1}&0&c_{m+2}A_{m+2}^{0,m-1}&\cdots&0&0\\
\vdots&\vdots&\vdots&\vdots&\ddots&\vdots&\vdots\\
0&0&0&0&\cdots&0&c_{\Lambda-1}A_{\Lambda-1}^{0,m-1}\\
0&0&0&0&\cdots&c_{\Lambda-1}A_{\Lambda-1}^{0,m-1}&0\\
\end{array}
\right).
\ee
{Lemma \ref{lemmadisugA}}, together with {proposition \ref{propoineqn2}}, (\ref{matr1}) and (\ref{matr2}), imply
$$
\|B_{m}\|_2<\|B_{m-1}^{n\left(\Lambda;m\right)}\|_2,
$$
so the proof is finished.
\endproof
\end{lemma}

At this point, let $\alpha_1\left(\Lambda\right):=\max{\left\{\alpha_1\left(\Lambda;0\right);\alpha_1\left(\Lambda;1\right);\cdots;\alpha_1\left(\Lambda;\Lambda\right) \right\}}$ and assume, per absurdum, that $\alpha_1\left(\Lambda\right)=\alpha_1\left(\Lambda;m\right)$ with $m>0$. We can take the matrix $B_{m-1}$ and its elements; from {lemma \ref{lemmadisugB}} we can infer
\be
\|B_{m}\|_2<\|B_{m-1}^{n\left(\Lambda;m\right)}\|_2;\label{disB1}
\ee
and from {lemma \ref{lemma1}} we know that the eigenvalues of
$B_{m-1}^{n\left(\Lambda;m\right)}$ ``separate'' the ones of $B_{m-1}$, then 
\be
\rho\left(B_{m-1}^{n\left(\Lambda;m\right)}\right)<\rho\left(B_{m-1}\right).
\label{disB2}
\ee
The inequalities (\ref{disB1}) and (\ref{disB2}) lead us to $\alpha_1\left(\Lambda\right)<\alpha_1\left(\Lambda;m-1\right)$, but this is not possible. We can then conclude that $\alpha_1\left(\Lambda\right)=\alpha_1\left(\Lambda;0\right)$ and with the same procedure we can prove the other inequalities in (\ref{disalpha}).
\subsubsection{Proof of (\ref{disalpha1}) in theorem \ref{teospec2}}
Let
$$
\widehat{B}_{0}\left(\Lambda\right):= \left(\!\!
\begin{array}{cccccccc}
0&\!A_{1}^{0,0}\!&0&0&0&0&0&0\\
A_{1}^{0,0}\!&0&\!A_{2}^{0,0}\!&0&0&0&0&0\\
0&\!A_{2}^{0,0}\!&0& \!A_{3}^{0,0}\!&0&0&0&0\\
\vdots&\vdots&\vdots&\vdots&\vdots&\vdots&\vdots&\vdots\\
0&0&0&0&0& A_{\Lambda-1}^{0,0}\! &0&A_{\Lambda}^{0,0}\\
0&0&0&0&0&0&A_{\Lambda}^{0,0}\! &0\\ 
\end{array}
\!\!\right)
$$
and its spectrum $\left\{\widehat{\alpha}_i\left(\Lambda;0\right) \right\}_{i=1}^{\Lambda+1}$, where the eigenvalues are arranged in descending order.

First of all, from $1\leq c_l\leq \sqrt{1+\frac{\Lambda^2}{k(\Lambda)}}$ $\forall 1\leq l\leq\Lambda$ and {proposition \ref{propoineqn2}} we obtain 
$$
\alpha_1\left(\Lambda;0\right)=\left\|B_0\left(\Lambda\right)\right\|_2\leq \sqrt{1+\frac{\Lambda^2}{k(\Lambda)}} \left\|\widehat{B}_0\left(\Lambda\right)\right\|_2 = \sqrt{1+\frac{\Lambda^2}{k(\Lambda)}}\widehat{\alpha}_1\left(\Lambda;0\right)
$$
and $\alpha_1\left(\Lambda+1;0\right)=\left\|B_0\left(\Lambda+1\right)\right\|_2\geq  \left\|\widehat{B}_0\left(\Lambda+1\right)\right\|_2 =\widehat{\alpha}_1\left(\Lambda+1;0\right)$; then, by algebraic calculations, one has
\be\label{alphahatineq}
\sqrt{1+\frac{\Lambda^2}{k(\Lambda)}} \widehat{\alpha}_1\left(\Lambda;0\right)\leq \widehat{\alpha}_1\left(\Lambda+1;0\right)\Leftrightarrow k(\Lambda)\geq \frac{\Lambda^2 \left[\widehat{\alpha}_1\left(\Lambda;0\right)\right]^2}{\left[\widehat{\alpha}_1\left(\Lambda+1;0\right)\right]^2-\left[\widehat{\alpha}_1\left(\Lambda;0\right)\right]^2}.
\ee

As done for {section \ref{proofteosimple1}}, one can use {theorem \ref{teoFavard}} and {theorem \ref{teodistinct}} to prove that $\widehat{\alpha}_1\left(\Lambda+1;0\right)> \widehat{\alpha}_1\left(\Lambda;0\right)$ $\forall\Lambda\in\mathbb{N}$, while it is obvious that
$$
\sqrt{\frac{l^2}{4l^2-1}}>\frac{1}{2}\quad \forall l\in\mathbb{N}\quad\Longrightarrow\quad \left\|\widehat{B}_0\left(\Lambda\right)\right\|_2=\widehat{\alpha}_1\left(\Lambda;0\right)>\cos{\left(\frac{\pi}{\Lambda+2} \right)}\quad\forall\Lambda\in\mathbb{N};
$$
finally, in {section \ref{proofdensity2}} we prove $\alpha_1\left(\Lambda;0\right)\rightarrow 1$ when $\Lambda\rightarrow+\infty$.

According to this, one has $\widehat{\alpha}_1\left(\Lambda;0\right)\uparrow 1$, $\widehat{\alpha}_1\left(\Lambda;0\right)= \cos{\left(\frac{\pi}{\Lambda+2} \right)}+\varepsilon(\Lambda)$ with $\varepsilon(\Lambda)\geq 0$ and $\varepsilon(\Lambda)\rightarrow 0$.

It is well known that $\cos{x}=1-\frac{x^2}{2}+o\left(x^3\right)$, then it is obvious that $\varepsilon(\Lambda)=\frac{1}{\Lambda}+o\left(\frac{1}{\Lambda} \right)$ when $\Lambda\rightarrow+\infty$ is not possible, because it is in constrast with $\widehat{\alpha}_1\left(\Lambda;0\right)= \cos{\left(\frac{\pi}{\Lambda+2} \right)}+\varepsilon(\Lambda)\leq 1$ $\forall\Lambda$; for the same reason, it must be
$$
\varepsilon(\Lambda)<\frac{\pi^2}{2\left(\Lambda+2\right)^2} \quad\mbox{when }\Lambda\rightarrow+\infty.
$$
Finally, this and 
$$
\cos{\left(\frac{\pi}{\Lambda+3} \right)}-\cos{\left(\frac{\pi}{\Lambda+2} \right)}=\frac{\pi}{\Lambda^3}+o\left(\frac{1}{\Lambda^3}\right)
$$
imply
$$
\widehat{\alpha}_1\left(\Lambda+1;0\right)-\widehat{\alpha}_1\left(\Lambda;0\right)=\frac{\widetilde{C}}{\Lambda^3}+o\left(\frac{1}{\Lambda^3}\right) \quad\mbox{when }\Lambda\rightarrow+\infty,
$$
for a suitable constant $\widetilde{C}>0$.

Coming back to (\ref{alphahatineq}), from $\sqrt{\frac{1}{3}}\leq\widehat{\alpha}_1\left(\Lambda;0\right)< 1$ $\forall\Lambda\in\mathbb{N}$ we can infer
$$
\frac{\Lambda^2 \left[\widehat{\alpha}_1\left(\Lambda;0\right)\right]^2}{\left[\widehat{\alpha}_1\left(\Lambda+1;0\right)\right]^2-\left[\widehat{\alpha}_1\left(\Lambda;0\right)\right]^2}\leq \frac{1}{2\sqrt{\frac{1}{3}}}\frac{\Lambda^2}{\widehat{\alpha}_1\left(\Lambda+1;0\right)-\widehat{\alpha}_1\left(\Lambda;0\right)}=\frac{1}{2\widetilde{C}\sqrt{\frac{1}{3}}}\Lambda^5+O\left(\Lambda^6\right)
$$
when $\Lambda\rightarrow+\infty$.

Then
$$
k(\Lambda)\geq \Lambda^6 \Rightarrow \alpha_1\left(\Lambda+1;0\right)>\alpha_1\left(\Lambda;0\right)\quad\mbox{definitively}.
$$
\subsubsection{Proof of item $(D)$ in theorem \ref{teospec2}}\label{proofdensity2}
First of all, from $\sqrt{\frac{l^2}{4l^2-1}}>\frac{1}{2}$ $\forall l\in\mathbb{N}$ and {proposition \ref{propoineqn2}} we obtain
\be\label{boundbasso}
\left\|P_{\Lambda+1}\left(0,\frac{1}{2},\frac{1}{2}\right)\right\|_2=\cos{\left(\frac{\pi}{\Lambda+2} \right)}< \alpha_1\left(\Lambda;0\right)=\left\|B_{0}\left(\Lambda\right)\right\|_2,
\ee
then the inequality (\ref{valuelimalpha2})$_2$ follows trivially from (\ref{boundbasso}), $\cos{x}\geq 1-\frac{x^2}{2}$ $\forall x\in[0,1]$ and $\frac{\pi}{\Lambda+2}\leq 1$ $\forall\Lambda\geq 2.$

On the other hand, if $\bm{\chi}_1$ is the $x_0$-eigenvector having $\alpha_1\left(\Lambda,0\right)$ eigenvalue, then $L_0 \bm{\chi}_1=0$, which implies
$$
\left\langle\bm{\chi}_1,x_+\bm{\chi}_1\right\rangle=0,\quad \left\langle\bm{\chi}_1,x_-\bm{\chi}_1\right\rangle=0\Rightarrow \left\langle\bm{\chi}_1,x_1\bm{\chi}_1\right\rangle=0,\quad \left\langle\bm{\chi}_1,x_2\bm{\chi}_1\right\rangle=0;
$$
so, from 
$$
\left(\Delta \bm{x}\right)^2_{\bm{\chi}_1}:= \left\langle \bm{\chi}_1,\bm{x}^2\bm{\chi}_1\right\rangle-\sum_{i=1}^3 \left\langle\bm{\chi}_1,x_i\bm{\chi}_1\right\rangle^2\geq0,
$$
we obtain 
\be\label{boundalto}
\left[\alpha_1\left(\Lambda,0\right)\right]^2=\left\langle \bm{\chi}_1,x_0\bm{\chi}_1\right\rangle^2\leq \left\langle \bm{\chi}_1,\bm{x}^2\bm{\chi}_1\right\rangle\overset{(\ref {defR2D=3})}\leq 1+\frac{\Lambda(\Lambda+1)+1}{k(\Lambda)}.
\ee
It is obvious that (\ref{boundbasso}) and (\ref{boundalto}) trivially imply
$$
\lim_{\Lambda\rightarrow+\infty}\alpha_1\left(\Lambda,0\right)=1.
$$
Once proved this, then the proof of $(D)$ is essentially the same of {section \ref{proofteodense1}}, the only difference is that here $A=P_{\Lambda+1}\left(0,\frac{1}{2},\frac{1}{2}\right)$, $A+E=B_0\left(\Lambda\right)$ and $\|E\|_2\leq 2\left\{\sqrt{1+\frac{1}{(\Lambda+1)^2}}\left[\frac{1}{2}+\frac{1}{12}\right]-\frac{1}{2}\right\} $, which follows from {proposition \ref{propoineqn2}}, (\ref{eigenvaluesToe}) and
\begin{equation*}
\begin{split}
c_l A_l^{0,m}&=\sqrt{1+\frac{l^2}{k(\Lambda)}}\sqrt{\frac{l^2-m^2}{4l^2-1}}\leq\sqrt{1+\frac{1}{(\Lambda+1)^2}}\sqrt{\frac{l^2}{4l^2-1}}\\
&\leq \sqrt{1+\frac{1}{(\Lambda+1)^2}}\left[\frac{1}{2}+\left(\sqrt{\frac{l^2}{4l^2-1}}-\frac{1}{2} \right) \right]\\
&=\sqrt{1+\frac{1}{(\Lambda+1)^2}}\left[\frac{1}{2}+\left(\frac{\frac{1}{4(4l^2-1)}}{\sqrt{\frac{l^2}{4l^2-1}}+\frac{1}{2}} \right) \right]\\
&\leq \sqrt{1+\frac{1}{(\Lambda+1)^2}}\left[\frac{1}{2}+\frac{1}{12}\right].
\end{split}
\end{equation*}


\begin{thebibliography}{9}
\bibitem{Madore} J. Madore, \emph{The Fuzzy sphere}, Classical and Quantum Gravity,  Volume 9,  Number 1, 1992.
\bibitem{HopdeWNic} 
J. Hoppe, \emph{Quantum theory of a massless relativistic surface and a two-dimensional bound state problem},  PhD thesis, MIT 1982; B. de Wit, J. Hoppe, H. Nicolai,
Nucl. Phys. {\bf B305} (1988), 545.
\bibitem{GroMad92}
H. Grosse, J. Madore, 
Phys. Lett. {\bf B283} (1992), 218.
\bibitem{GroKliPre96'}
H. Grosse, C. Klimcik, P. Presnajder,
Int. J. Theor. Phys. {\bf 35} (1996), 231-244.
\bibitem{GroKliPre96}
H. Grosse, C. Klimcik, P. Presnajder,
	Commun. Math. Phys. {\bf 180} (1996), 429-438.
\bibitem{Ramgoolam}
S. Ramgoolam,
Nucl. Phys. {\bf B610} (2001), 461-488
;
  JHEP10(2002)064;
and references therein.
\bibitem{Dolan:2003th}
 B. P. Dolan, D. O'Connor,  P. Presnajder,    {\it Fuzzy complex quadrics and spheres},
 JHEP {\bf 0402} (2004) 055; and references therein.
\bibitem{Ste17}
M. Sperling, H. Steinacker,  
J. Phys. A: Math. Theor. {\bf 50} (2017), 375202
\bibitem{AscMadManSteZou}
P. Aschieri,  H. Steinacker, J. Madore,   P. Manousselis, G. Zoupanos
SFIN {\bf A1} (2007) 25-42;
and references therein.
\bibitem{AlessioArzano} F. Alessio, M. Arzano, \emph{A fuzzy bipolar celestial sphere}, J. High Energ. Phys., {\bf 28} (2019), doi:10.1007/JHEP07(2019)028.
\bibitem{Banks} T. Banks,W. Fischler, S. H. Shenker and L. Susskind, \emph{M Theory As A Matrix Model: A Conjecture}, Phys.Rev.D55:5112-5128, 1997.
\bibitem{Berko} M. Berkooz, M. R. Douglas, \emph{Five-branes in M(atrix) Theory}, Phys.Lett. B395 (1997) 196-202.
\bibitem{IKKT97} N. Ishibashi, H. Kawai, Y. Kitazawa,  A. Tsuchiya, {\it A Large $N$ reduced model as superstring}, Nucl. Phys. B498 (1997), 467491. 
\bibitem{FiorePisacaneJGP18}
G.Fiore, F.Pisacane, 
{\it Fuzzy circle and new fuzzy sphere through confining potentials and energy cutoffs},
J. Geom. Phys. {\bf 132} (2018), 423-451.
\bibitem{FiorePisacanePOS18}
G.Fiore, F.Pisacane, 
{\it New fuzzy spheres through confining potentials and energy cutoffs}, 
Proceedings of Science Volume 318,
PoS(CORFU2017)184.
\bibitem{Snyder}
H. S. Snyder, \emph{Quantized Space-Time},
Phys. Rev. {\bf 71} (1947), 38.
\bibitem{CS-FiorePisacane}
G. Fiore, F. Pisacane, 
{\it On localized and coherent states on some new fuzzy spheres}, 
arXiv:1906.01881.
\bibitem{BarrettEtAl2011}
J.W. Barrett, 
et al.,
Gen. Relativ. Gravit. {\bf 43} (2011), 2421. 
\bibitem{Ste16NPB}
H. Steinacker,
Nucl. Phys. {\bf B910} (2016),  346-373
\bibitem{Connes} A. Connes,
\emph{Noncommutative geometry},
Academic Press, 1995.
\bibitem{ConCha10} A. H. Chamseddine, A. Connes. {\it Space-time from the spectral point of view}, Proc.
of the 12th Marcel Grossmann meeting, World Scientific, 2012, pp. 3-23;
and references therein.
\bibitem{DanLizMar14} 
F. D’Andrea, F. Lizzi, P. Martinetti,
\emph{Spectral geometry with a cut-off: Topological and metric aspects},
J. Geom. Phys.  {\bf  82} (2014), 18-45.
\bibitem{BahDopFrePia11} D. Bahns, S. Doplicher, K. Fredenhagen,  G. Piacitelli, {\it Quantum geometry on quantum spacetime: distance, area and volume operators}, Commun. Math. Phys., {\bf  308} (2011), 567-589.
\bibitem{MarMerTom12}
P. Martinetti, F. Mercati, L. Tomassini
{\it Minimal length in quantum space and integrations of the line element in Noncommutative Geometry},
Rev. Math. Phys. {\bf  24}  (2012), 1250010
\bibitem{Noschese} S. Noschese, L. Pasquini, L. Reichel \emph{Tridiagonal Toeplitz matrices: properties and novel applications}, Numerical Linear Algebra with applications, {\bf  20},  2013.
\bibitem{Freud} G. Freud,
\emph{Orthogonal Polynomials}, Pergamon Press, 1966.
\bibitem{Szego} G. Szego,
\emph{Orthogonal Polynomials}, American Mathematical Society Colloquium Pubblications, XXIII, 1939.
\bibitem{Horn} R. A. Horn, C. R. Johnson
\emph{Matrix Analysis}, Cambridge University Press, 1990.
\end{thebibliography}
\end{document}